\documentclass[11pt,a4paper]{article}
\pdfoutput=1
\usepackage{jheppub}
\usepackage[T1]{fontenc}
\usepackage[utf8]{inputenc}
\usepackage{slashed}
\usepackage{bm}
\usepackage[bottom]{footmisc}
\newcommand{\be}{\begin{equation}}
\newcommand{\ee}{\end{equation}}
\newcommand{\M}{\mathcal{M}}
\newcommand{\dd}{\textmd{d}}
\newcommand{\Z}{\mathcal{Z}}
\newcommand{\D}{\mathcal{D}}
\newcommand{\expv}[1]{\left \langle #1 \right \rangle}
\newcommand{\tr}{\textmd{tr}\,}
\newcommand{\dsf}{\slashed{D}_f}
\newcommand{\subs}{{12}}
\newcommand{\muqed}{\mu_{\rm QED}}
\newcommand{\muqcd}{\mu_{\rm QCD}}
\newcommand{\Regensburg}{Institut f\"ur Theoretische Physik, Universit\"at Regensburg, D-93040 Regensburg, Germany.}
\newcommand{\Tata}{Tata Institute of Fundamental Research, Homi Bhabha Road, Mumbai 400005, India.}
\newcommand{\Frankfurt}{Institut f\"ur Theoretische Physik, Goethe Universit\"at Frankfurt, D-60438 Frankfurt am Main, Germany.}
\newcommand{\Bielefeld}{Fakult\"at f\"ur Physik, Universit\"at Bielefeld,
  D-33615 Bielefeld, Germany.}

\title{
Magnetic susceptibility of QCD matter and its decomposition from the lattice
}
\author[a,b]{Gunnar.~S.~Bali,}
\author[c,d]{Gergely~Endr\H{o}di,}
\author[a]{and Stefano~Piemonte}
\affiliation[a]{\Regensburg}
\affiliation[b]{\Tata}
\affiliation[c]{\Frankfurt}
\affiliation[d]{\Bielefeld}
\emailAdd{gunnar.bali@ur.de}
\emailAdd{endrodi@physik.uni-bielefeld.de}
\emailAdd{stefano.piemonte@ur.de}

\abstract{
We determine the magnetic susceptibility of thermal QCD matter by means of first principles lattice 
simulations using staggered quarks with physical masses.
A novel method is employed that only requires simulations 
at zero background field, thereby circumventing problems related to magnetic flux quantization. After a careful
continuum limit extrapolation, diamagnetic behavior (negative susceptibility) is found at low temperatures
and strong paramagnetism (positive susceptibility) at high temperatures. We revisit the decomposition of the 
magnetic susceptibility into spin- and orbital angular momentum-related contributions. The spin term -- related to the normalization
of the photon lightcone distribution amplitude at zero temperature --
is calculated non-perturbatively and extrapolated to the continuum
limit.
Having access to both the full magnetic susceptibility and 
the spin term, we calculate the orbital angular momentum contribution
for the first time. The results reveal the opposite of what might be 
expected based on a free fermion picture.
We provide a simple parametrization of the temperature- and magnetic
field-dependence of the QCD equation of state
that can be used in phenomenological studies.}

\begin{document}
\maketitle
\section{Introduction}
\label{sec:intro}
The development of a quantitative and precise understanding of the response
of QCD matter to background (electro)magnetic fields is
of vital importance for furthering our knowledge about a multitude
of physical systems. Examples include the
interior of magnetars, neutron star mergers~\cite{Kiuchi:2015sga,Baiotti:2016qnr,Kawamura:2016nmk}, off-central
heavy-ion collisions and 
the evolution of the universe in its early stages. For general reviews, we refer
the reader to Refs.~\cite{Kharzeev:2013jha,Andersen:2014xxa,Miransky:2015ava}. 
A characteristic feature of the behavior of strongly interacting quarks and 
gluons is rooted in the dependence of the QCD equation of state (EoS) on the background
magnetic field $B$. The EoS enters all
the above mentioned 
examples: it appears in the gravitational stability conditions 
of compact stars, affecting the mass-radius relation~\cite{Lattimer:2000nx};
it governs the expansion rate in cosmological
models~\cite{Grasso:2000wj,Durrer:2013pga} and it also 
sets the conditions where freeze-out is reached in heavy-ion collisions,
see, e.g., Ref.~\cite{Kharzeev:2015znc}.

While the equilibration of fireballs produced in heavy-ion collisions
is still a subject of research, at least in astrophysical systems
the time and distance scales over which the magnetic field varies are much
larger than those that govern QCD processes that affect, e.g., the EoS
or nucleo-synthesis. With these applications
in mind, solving QCD in a constant background magnetic field
is sufficient and this scenario is amenable to lattice simulations.

The leading dependence of the EoS on $B$ is encoded in the magnetic
susceptibility  $\chi$ of QCD matter. Its sign distinguishes between
paramagnets ($\chi>0$), for which the exposure to the background field is
energetically favorable, and 
diamagnets ($\chi<0$), which repel the external field.
In QCD matter, like in any other material, the origin of the magnetization
and hence of the magnetic susceptibility is related to the spin and
angular momentum of charged particles. At high temperatures the quarks are the relevant degrees of freedom; at low temperatures the hadrons and in particular the pions take over, while (valence and sea) quarks contribute just as their fundamental constituents.

The total angular momentum
that gives rise to the magnetization can be decomposed into contributions
from the spins of the quarks of different flavors
and a remainder. The latter contains
the quark orbital angular momenta but also the angular momentum
of the gluons, that can split into quark-antiquark pairs.\footnote{This
  situation is analogous to the decomposition of the nucleon spin.
  In particular, vacuum expectation values of the same local operators
  appear in the magnetic field background at zero momentum
  as in the decomposition~\cite{Ji:1996ek} of the transversely polarized
  generalized parton distribution functions of deep inelastic scattering
  at leading twist. While the individual quark spin contributions in both
  cases are unique and gauge invariant, the further decomposition of the
  remainder into quark and gluon parts is ambiguous: the decomposition
  of Ref.~\cite{Ji:1996ek} is based on the Belinfante-Rosenfeld form of
  the energy momentum tensor, which is also the natural starting point
  for lattice QCD, but one may also, e.g., resort to the canonical
  definition of the angular momentum~\cite{Jaffe:1989jz,Bakker:2004ib}.}
The quark spin contribution to the magnetic susceptibility is due
to the expectation value $\expv{\bar\psi_f\sigma_{\mu\nu}\psi_f}$, whose leading
order response is linear in the magnetic field strength tensor 
$F_{\mu\nu}$. Depending on the normalization, the
slope is proportional to the
magnetic susceptibility of the quark
condensate~\cite{Ioffe:1983ju,Belyaev:1984ic,Balitsky:1985aq}, or the
so-called tensor coefficient, i.e.\ the normalization of the photon
lightcone distribution
amplitude (DA)~\cite{Balitsky:1989ry,Petrov:1998kg,Ball:2002ps} of finding a quark-antiquark
pair of flavor $f$ in a transverse photon. This in itself appears in a multitude
of applications, e.g., as a correction to the hadronic light-by-light
scattering contribution to the muon anomalous magnetic
moment~\cite{Nyffeler:2009tw,Bijnens:2019ghy},
within radiative transitions~\cite{Ball:2003fq,Colangelo:2005hv} and in the
photo-production of mesons~\cite{Agaev:2014wna}.

Some of the present authors have already addressed several of the
above aspects in Refs.~\cite{Bali:2012zg,Bali:2012jv,Bali:2013esa,Bali:2014kia}.
Here we improve on these studies by employing a novel calculational method
that is based on Ref.~\cite{Bali:2015msa}, by carrying out the QCD
renormalization non-perturbatively with respect to the intermediate
RI'-MOM scheme~\cite{Martinelli:1994ty,Chetyrkin:1999pq} and by adding
a finer lattice spacing. In addition we present and exploit new analytical
findings. One of the outcomes will be that in the strongly interacting medium
at low to moderately high temperatures the quark spin-related
susceptibility is negative (diamagnetism) while the part that is
due to the orbital angular momentum is positive. Clearly,
this behavior
is very different from the response to magnetic
fields of the materials that have so far been accessible to
solid state physics experiments.
Therefore, our results offer a glimpse into a completely new
regime of spin physics.

The article is organized as follows.
In Sec.~\ref{sec:response} we introduce our notations and the
central observables. For conceptual clarity, we carefully address their
divergence structure and renormalization in QED and QCD.
In Sec.~\ref{sec:latsetup} we then
discuss details of the simulation and, in particular, we introduce
our new method that employs current-current correlators in a mixed
coordinate- and momentum-space representation. This enables us to
determine susceptibilities from lattice simulations at $B=0$.
We then present and discuss our results in Sec.~\ref{sec:res}, before
we summarize. We include several technical appendices: in App.~\ref{sec:B}
we investigate the effects of taste splitting in the staggered formulation
within the hadron resonance gas model. This turns out to be
important to avoid
underestimating the systematics of the continuum limit extrapolation.
In App.~\ref{app:spinpart} we derive the factorization of the susceptibility
into quark spin-related and other contributions, building
upon earlier partial results~\cite{Bali:2012jv}. In the extensive
App.~\ref{app:freecase}, several derivations are carried out for
the free case, establishing, e.g., the structure of QED divergencies.
App.~\ref{sec:renormconstants} discusses the non-perturbative
renormalization procedure. In App.~\ref{app:derivatives}
we present more detail on the derivation of the
new current-current method and compare
to numerical results, obtained using conventional background field
approaches.
Finally, App.~\ref{app:parameterization} gives a parametrization of our results for 
the QCD EoS 
for a broad range of temperatures and magnetic field strengths.
The corresponding Python script \texttt{param\_EoS.py} is uploaded to the arXiv as ancillary file along with this paper.

\section{The response of QCD matter to background fields and the magnetic susceptibility}
\label{sec:response}
Without any loss of generality, below we consider a magnetic field pointing 
in the $x_3$ direction, with the magnitude $B$. 
The magnetic susceptibility is defined via the leading (quadratic) dependence of 
the QCD free energy density $f$ on the field strength,
\be
\chi_b= -\left.\frac{\partial^2 f}{\partial (eB)^2}\right|_{B=0}, \quad\quad
f = -\frac{T}{V}\log\Z\,,
\label{eq:defchi}
\ee
where $\Z$ is the partition function, $T$ the temperature and $V$ the spatial volume of the system. 
The product $eB$ of the elementary electric charge $e$ and the magnetic field 
is a renormalization group invariant due to the QED vector Ward identity, see Eq.~\eqref{eq:eBinv} below. Therefore, $\chi_b$ is free of multiplicative 
renormalization.
Still, the susceptibility undergoes additive renormalization, which is made explicit by the index $b$,
indicating the bare quantity, as obtained in the lattice scheme at
  a lattice spacing $a$.\footnote{Below we will indicate quantities that
  are subject to QED renormalization with the subscript $b$.}
This was discussed in Ref.~\cite{Bali:2014kia} in depth, but we repeat the argument here 
for comprehensiveness.

\subsection{QED renormalization}

The total free energy density
of the system, that includes both QCD matter and the classical background field,
\be
f_{\rm tot} = f + \frac{B_b^2}{2}\,,
\label{eq:totalf}
\ee
is a physical observable and therefore free of divergences.
However, the second term within Eq.~\eqref{eq:totalf}, involving 
the bare magnetic field $B_b$, contains a logarithmic divergence in the lattice spacing $a$ due to 
electric charge renormalization~\cite{Schwinger:1951nm},
\be
B_b^2=Z_e B^2, \quad\quad e_b^2=Z_e^{-1}e^2, \quad\quad 
eB = e_bB_b, \quad\quad Z_e=1+\beta_1(a^{-1}) \, e^2\,\log\left(\muqed^2 a^2\right) \,,
\label{eq:eBinv}
\ee
where the renormalized quantities $e$ and $B$ depend on the QED renormalization
scheme and on the QED renormalization scale $\muqed$.
Since the background field is classical, only the leading-order QED $\beta$-function coefficient $\beta_1$ appears here~\cite{Dunne:2004nc}. Note that $\beta_1$ is affected by QCD corrections at the 
cut-off scale,
\be
\beta_1(a^{-1})=\beta_1 \cdot \Bigg[ 1 + \sum_{i\ge 1} c_i \,g^{2i}(a^{-1})\Bigg]\xrightarrow{a\to0}\beta_1, \quad\quad\beta_1=\frac{1}{4\pi^2} \cdot \sum_f (q_f/e)^2\,,
\label{eq:beta1def}
\ee
where $g$ is the strong coupling and 
$q_f$ denotes the electric charge of the quark flavor $f$. The coefficients $c_i$ of the perturbative series are known up to $i=4$ in the $\overline{\mathrm{MS}}$
and $\mathrm{MOM}$ schemes~\cite{Baikov:2012zm} and $c_1=1/(4\pi^2)$
is universal for massless schemes.
These QCD corrections\footnote{Note that in general disconnected diagrams 
also start to contribute for $i\geq 3$, but in the present case 
these vanish because we 
are dealing with the three lightest quark flavors and $\sum_f q_f/e=0$.}
vanish logarithmically with the lattice spacing towards
the continuum limit due to the asymptotically free nature of the
strong interactions, as is also
indicated in Eq.~\eqref{eq:beta1def}.

Eq.~\eqref{eq:totalf} implies that $f$ contains the same additive divergence as $B_b^2/2$, but with an opposite sign. This propagates into the 
susceptibility~\eqref{eq:defchi}, resulting in
\be
\chi_b = \chi[\muqed] + \beta_1(a^{-1})\, \log (\muqed^2a^2)\,.
\label{eq:b1}
\ee
The renormalized susceptibility $\chi$ depends on the renormalization 
scale, which we indicated here explicitly in square brackets.
We confirm the presence of the logarithmic divergence in $\chi_b$ 
analytically for the free case in App.~\ref{sec:UVdivs} and numerically 
for full QCD in Sec.~\ref{sec:res_1}. 
The divergence is independent of the temperature so that it cancels within the difference
\be
\chi \equiv \chi[\muqed^{\rm phys}]= \chi_b(T) - \chi_b(T=0)\,.
\label{eq:renorm}
\ee
This definition of the renormalized susceptibility -- implying that it vanishes 
identically at $T=0$ -- 
corresponds to a particular choice of the renormalization scale $\muqed=\muqed^{\rm phys}$. In fact this is the only prescription that adheres to the
physical requirement that the magnetic permeability $(1-e^2\chi)^{-1}$
should be unity in the vacuum.
In the following we suppress the dependence on the QED renormalization
scale and simply write $\chi$ for the susceptibility, renormalized
in this way.

\subsection{The tensor coefficient}

Besides the magnetic susceptibility there exist further 
quantities that characterize the leading-order response 
of the QCD medium to the background magnetic field.
For a general background field $F_{\mu\nu}$, 
the fermion bilinear involving the 
relativistic spin operator $\sigma_{\mu\nu}$ develops a nonzero
expectation value~\cite{Ioffe:1983ju,Balitsky:1983xk},
\be
\expv{\bar\psi_f \sigma_{\mu\nu} \psi_f}
= q_f F_{\mu\nu} \cdot \tau_{fb} + \mathcal{O}(F^3)\,, 
\quad\quad 
\sigma_{\mu\nu} = \frac{1}{2i}[\gamma_\mu, \gamma_\nu]\,.
\label{eq:sxydef}
\ee
We will refer to $\tau_{fb}$ as the tensor coefficient for the 
flavor $f$. Similarly to $\chi_b$, this is also a bare observable 
that contains additive logarithmic divergences in the cut-off.
For our choice of direction of the magnetic field
the tensor coefficient can be determined as the slope of
the expectation value of the fermion bilinear
involving $\sigma_\subs$ at small values of the magnetic field:
\be
\tau_{fb} = \lim_{B\to0}\frac{\expv{\bar\psi_f\sigma_\subs\psi_f}}{q_fB}\,.
\label{eq:taufbdefinition}
\ee

The tensor coefficient contains a similar logarithmic divergence as $\chi_b$. We demonstrate 
the reason for this in Sec.~\ref{sec:decomp1} below.
In particular, the divergence structure takes the form~\cite{Bali:2012jv},
\be
\tau_{fb}= \tau_f + \gamma_1^{\tau}(a^{-1})\, m_{f} \log(\muqed^2 a^2)\,, \quad\quad 
\gamma_{1}^{\tau}(a^{-1}) = \gamma^{\tau}_1\cdot \left[ 1 + \mathcal{O}(g^2(a^{-1}))\right]\,, \quad\;
\gamma_1^\tau = \frac{3}{4\pi^2}\,,
\label{eq:t1}
\ee
where $m_f$ is the mass of the quark of flavor $f$.
Again, due to asymptotic freedom, QCD corrections to $\gamma_1^\tau$ vanish in the 
continuum limit.\footnote{Unlike for $\beta_1(a^{-1})$ in Eq.~\eqref{eq:beta1def}, the order $g^2$ perturbative coefficient is not known in this case. 
  Therefore, any definition of a renormalized
  quark mass $m_f$ is valid to this order and we use the lattice quark mass.}
Eq.~\eqref{eq:t1} is confirmed in App.~\ref{sec:tensorbfree} for the free case and 
checked numerically in full QCD in Sec.~\ref{sec:res_2}.
Notice that in the chiral limit the divergent term disappears, so 
that $\lim_{m_f\to0} \tau_{fb}$ is ultraviolet-finite. We will
carry out this limit at 
zero temperature in Sec.~\ref{sec:res_2} below.
In this situation, up to multiplicative renormalization,
  this object corresponds to the normalization
  $f_{\gamma}^{\perp}$ of the leading-twist photon distribution
  amplitude~\cite{Balitsky:1989ry,Petrov:1998kg,Ball:2002ps},
  i.e.\ of the infinite momentum
  frame probability amplitude
  that a real photon dissociates into a quark-antiquark pair of
  flavor $f$.

In analogy to Eq.~\eqref{eq:renorm}, we define the renormalized tensor
coefficient by subtracting its value at zero temperature,
\be
\tau_f = \tau_{fb}(T) - \tau_{fb}(T=0)\,,
\label{eq:taufTsub}
\ee
which again corresponds to a particular choice of the QED renormalization scale. 
Unlike for $\chi$, there is no preferred choice in this case, but it is natural to 
use the same prescription as in Eq.~\eqref{eq:renorm} above.

Besides the (QED-related) additive renormalization detailed above, the tensor coefficient also 
undergoes (QCD-related) 
multiplicative renormalization by the tensor renormalization constant $Z_T$. 
This introduces a further scheme- and scale-dependence of this observable. Below we 
will consider $Z_T$ in the $\overline{\rm MS}$ scheme at the QCD renormalization 
scale $\muqcd=2\textmd{ GeV}$. Unlike in our previous
study~\cite{Bali:2012jv}, 
where we calculated $Z_T$ perturbatively at the one-loop level,
  here we carry out a non-perturbative matching to
  to the RI'-MOM scheme~\cite{Martinelli:1994ty,Chetyrkin:1999pq}
  and subsequently translate the result at three-loop order~\cite{Gracey:2003yr}
  into the $\overline{\rm MS}$ scheme.
This procedure is detailed in App.~\ref{sec:renormconstants}.
We remark that $Z_T$ is independent of the temperature,
thus the ordering of the QED renormalization (i.e.\ the $T=0$ subtraction)
and the QCD renormalization 
(multiplication by $Z_T$) is irrelevant for the determination of
the renormalized tensor coefficient $\tau_f$.

\subsection{Decomposition into spin and orbital angular momentum contributions}
\label{sec:decomp1}

One might suspect that $\chi$ and $\tau_f$ are not completely unrelated. 
Indeed, as we first discussed in Ref.~\cite{Bali:2012jv}, 
$\tau_{f}$ represents the contribution of the spin of the quark flavor $f$ to the total magnetic susceptibility. 
In particular, $\chi$ can be decomposed into spin-related and orbital angular momentum-related contributions,\footnote{Note that for simplicity we refer to $\chi^{\rm ang}$ as the orbital
  angular momentum contribution. In the interacting case this can be
  further factorized, separating out the gluon total angular momentum
  contribution~\cite{Bali:2013esa} from those of the quark angular momenta.}
\be
\chi = \chi^{\rm spin} + \chi^{\rm ang}\,,
\label{eq:separation}
\ee
and the spin term is related to the
tensor coefficients as
\be
\chi^{\rm spin} = \sum_f \frac{(q_f/e)^2}{2m_{f}} \left[ \tau_{f}(m_f^{\rm val}) - \tau_{f}(m_f^{\rm val}\to0) \right ]\cdot Z_T Z_S \,,
\label{eq:spinpart1}
\ee
where $m_f$ denotes the quark mass in the lattice scheme and $Z_S$ and
$Z_T$ are the scalar and tensor renormalization constants, respectively.
The second term in the square brackets is understood
to correspond to the limit of a vanishing valence quark mass
taken at physical, i.e.\ nonzero, values of the sea quark masses.
We discuss the difference between valence 
and sea quark masses in Sec.~\ref{sec:latsetup} below.
In App.~\ref{app:spinpart} we prove Eqs.~\eqref{eq:separation}--\eqref{eq:spinpart1} and illustrate the origin of the 
subtraction of the chiral valence quark limit.
App.~\ref{app:freecase} also contains an explicit check of Eq.~\eqref{eq:spinpart1} in the free case.

A remark regarding the choice of renormalization scales is in order here.
Eq.~\eqref{eq:spinpart1} is chosen so that $\chi^{\rm spin}$ vanishes at $T=0$. 
Recall however that, according to our remark below Eq.~\eqref{eq:taufTsub}, 
we are free to choose an arbitrary QED renormalization scale for 
the renormalized tensor coefficient. 
This freedom propagates into $\chi^{\rm spin}$ and -- through the
decomposition~\eqref{eq:separation} -- 
to $\chi^{\rm ang}$ as well. 
In contrast, the renormalization scale for $\chi$ 
is fixed by the requirement $\chi(T=0)=0$. 
Thus, in principle both susceptibility 
contributions may be shifted by an arbitrary amount, as long as
their sum remains zero at $T=0$.
We will follow the choice made in Eq.~\eqref{eq:spinpart1},
which corresponds to setting $\chi^{\rm spin}(T=0)=\chi^{\rm ang}(T=0)=0$.
In the free case this is realized by choosing {\it one and the same}\
QED renormalization scale for all susceptibility contributions,
see App.~\ref{sec:spincontribapp}. Our numerical results in full QCD
  below suggest that also in the interacting case
  the QED scale 
  that corresponds to the renormalization
  condition $\chi^{\rm spin}(T=0)=0$ is consistent with the one obtained
  from setting $\chi(T=0)=0$.

In Eq.~\eqref{eq:spinpart1} we 
also carried out the QCD related multiplicative renormalization 
by including the tensor renormalization constant $Z_T$ 
required for $\tau_f$ (as mentioned above)
as well as the scalar renormalization constant $Z_S=Z_m^{-1}$, 
which multiplies the inverse quark mass.\footnote{Like in massless continuum
  schemes, in staggered lattice formulations there is no difference
  between singlet and non-singlet
  renormalization factors for these quark bilinears. This was
  explicitly demonstrated at order $g^4$ in
  Ref.~\cite{Constantinou:2016ieh}.} Note that these 
renormalization factors depend on the QCD regularization scheme and
on the QCD renormalization 
scale. As mentioned above, we choose the $\overline{\mathrm{MS}}$ scheme and $\muqcd=2\textmd{ GeV}$. We remark that, just as for $\tau_f$, the ordering 
of the QED and the QCD renormalization is irrelevant for $\chi^{\rm spin}$.
We stress again that while the factorization of the total susceptibility $\chi$ into
  $\chi^{\rm spin}$ and $\chi^{\rm ang}$
  depends on the QCD scheme and scale, $\chi$ itself is a
  QCD renormalization group invariant.

In the free case the two susceptibility contributions have a constant ratio,
$\chi^{\rm spin}:\chi^{\rm ang}=3:(-1)$, reflecting the well-known 
response 
of a free charged fermion to the magnetic field via its 
spin and its orbital angular momentum, dating back to Pauli and Landau~\cite{Pauli:1927:GPG,Landau1930}. 
This ratio, which translates into the rule $\chi^{\rm spin}:\chi=3:2$, holds identically in the free case, see 
App.~\ref{sec:spincontribapp}. In contrast, in full QCD it only applies 
to the divergence structure, which in the continuum limit
-- as we have seen above -- 
is governed by pure QED physics.
Below we determine to what extent the Pauli-Landau decomposition 
is affected by the strong interactions.

\section{Lattice methods}
\label{sec:latsetup}

We consider spatially symmetric $N_s^3\times N_t$ lattice ensembles,
corresponding to the temperature $T=1/(N_ta)$ and 
the volume $V=L^3=(N_sa)^3$. The simulations are performed with the 
tree-level Symanzik improved gauge action $S_g$ 
and three flavors ($f=u,d,s$) of 
stout smeared rooted staggered quarks~\cite{Aoki:2005vt}, 
described by the Dirac operator\footnote{Due to the electromagnetic
      charge $q_f$, the covariant derivative depends on the
      quark flavor $f$.
  } $\dsf+m_f$. 
The quark masses $m_f$ are tuned as a function of the inverse gauge coupling $\beta=6/g^2$ along the line of constant physics: $m_{ud}(\beta)\equiv m_u(\beta)=m_d(\beta)=m_s(\beta)/R$ 
with $R=28.15$~\cite{Borsanyi:2010cj}. 
The electric charges are set as $q_d=q_s=-q_u/2=-e/3$, where 
$e>0$ is the elementary electric charge. The magnetic 
field enters in $\dsf$ via space-dependent $\mathrm{U}(1)$ phases. Further details of 
our setup and of the simulation algorithm are discussed in Ref.~\cite{Bali:2011qj}.
The lattice geometries for our finite temperature lattices are $16^3\times 6$, 
$24^3\times 6$, $24^3\times 8$, $28^3\times10$ and $36^3\times12$, allowing 
for the investigation of both finite volume and discretization effects.
Our zero-temperature ensembles consist of $24^3\times32$, $32^3\times48$ and 
$40^3\times 48$ lattices.

In the rooted staggered formulation the partition function and 
the expectation value of the tensor bilinear are written as path 
integrals over the $\mathrm{SU}(3)$-valued gluonic links $U$ as
\be
\begin{split}
\Z &= \int \D U\, e^{-\beta S_g} \prod_{f'=u,d,s} \left[\det (\slashed{D}_{f'}+m_{f'}^{\rm sea})\right]^{1/4}\,,\\
\expv{\bar\psi_f\sigma_\subs\psi_f} &= 
\frac{T}{V} \frac{1}{\Z} \int \D U \, e^{-\beta S_g}  
\;\tr \frac{\sigma_\subs}{\dsf+m_f^{\rm val}}
\prod_{f'=u,d,s} \left[\det (\slashed{D}_{f'}+m_{f'}^{\rm sea})\right]^{1/4}\,.\\
\end{split}
\label{eq:Zseaval}
\ee
Here we distinguished between two different types of masses: the sea quark 
masses $m_f^{\rm sea}$ which appear in the fermion determinant and thus 
affect the generation of gluonic configurations; and the valence 
quark mass $m_f^{\rm val}$, which enters in the operator and thereby 
affects the measurement on a given set of configurations.
For usual observables both masses are equal and set according to the line 
of constant physics, $m_f^{\rm sea}=m_f^{\rm val}=m_f$.
The spin contribution to the magnetic susceptibility is exceptional 
in this sense -- 
as pointed out above in Eq.~\eqref{eq:spinpart1}, 
it also involves the value of $\expv{\bar\psi_f\sigma_\subs\psi_f}$ 
in the limit $m_f^{\rm val}\to0$ but keeping $m_f^{\rm sea}=m_f$.
Note that this does not mean that we are dealing with a non-unitary theory
but merely 
that $\chi^{\rm spin}$ can be expressed as a difference of two expectation 
values involving $\tau_f$ at different valence quark mass values,
see App.~\ref{app:spinpart}.

\subsection{Magnetic flux quantization}
\label{sec:fluxquant}

In an infinite volume a magnetic field pointing in the $x_3$ direction 
can be generated by the Landau-gauge electromagnetic potential 
\be
A_2=B x_1\,.
\label{eq:gaugechoice}
\ee
In a finite volume, in order to comply with periodic boundary 
conditions for the electromagnetic parallel transporters
  $u_{\mu f}=\exp (i q_fA_\mu)$,
the boundary twist term $A_1=-B x_2 L \, \delta(x_1-L)$
needs to be included as well~\cite{Martinelli:1982cb}.
In this setup the flux of the magnetic field is quantized according to~\cite{AlHashimi:2008hr}
\be
eB = 6\pi N_B / L^2, \quad\quad N_B\in\mathbb{Z}\,,
\ee
so that a differentiation with respect to $eB$ -- as required in Eq.~\eqref{eq:defchi} --  is not possible in a standard manner. 
Several methods were developed to overcome this problem on the lattice, including 
the anisotropy method~\cite{Bali:2013esa,Bali:2013owa}, the finite difference method~\cite{Bonati:2013lca,Bonati:2013vba} 
and the generalized integral method~\cite{Bali:2014kia}. These are all based on approximating 
the derivative numerically using finite differences in the integer variable $N_B$. This 
requires independent simulations using several values of $N_B$, which increases the computational requirements 
considerably. Furthermore, an extrapolation $N_B\to0$ becomes necessary, 
which inevitably introduces systematic uncertainties.
An alternative approach is the half-half method~\cite{Levkova:2013qda}, which 
employs a magnetic 
field profile that is positive in one half and negative in the other half of the lattice -- 
this enables taking the derivative with respect to the amplitude of the field analytically.
However, finite volume effects are substantially enhanced in this case due to the discontinuity 
of the background field at the boundaries~\cite{Bali:2015msa} -- 
even if such effects are expected to cancel in temperature 
differences~\cite{Ludmilla}.

\subsection{Determining the susceptibility via current-current correlators}

In view of the above, a method is desirable that only involves
measurements at $B=0$, thereby circumventing the flux quantization
problem of the constant background field profile.
In Ref.~\cite{Bali:2015msa} we demonstrated that
at zero temperature, 
$\chi_b$, as defined in \eqref{eq:defchi},
is related to a mixed-representation two-point 
function of the electromagnetic current, and can thus be 
measured at $B=0$. Below we motivate this method and 
clarify how it extends to nonzero temperatures.
A detailed derivation can be found in App.~\ref{app:derivatives} and,
for $T=0$, in Ref.~\cite{Bali:2015msa}.

Before integrating out the fermions in the path integral~\eqref{eq:Zseaval}, the vector potential~\eqref{eq:gaugechoice} couples to $i\cdot e$ times
the $\mu=2$ component of the electromagnetic current,
\begin{equation}
\label{eq:current}
j_{\mu}=\sum_f \frac{q_f}{e}\, \bar\psi_f \gamma_\mu \psi_f
\,.
\end{equation}
in the action density. Taking derivatives of $\log\Z$ with 
respect to $eB$ therefore brings down integrals 
over the current $j_2$ times the $x_1$-dependent term  
$i\,\partial A_2/\partial B$. Thus we can anticipate the result to take the form 
of a convolution of the projected correlator, 
\be
G(x_1) = \int \dd x_2 \,\dd x_3 \,\dd x_4 \,\expv{j_2(x)j_2(0)} \,,
\label{eq:G22def}
\ee
with an $x_1$-dependent kernel. 

Instead of directly using the gauge~\eqref{eq:gaugechoice}, 
it is instructive to approach the constant magnetic field 
background via oscillatory fields that possess nonzero 
momentum $p_1$ in the $x_1$ direction. Using these profiles, 
we can take
the thermodynamic limit and subsequently the $p_1\to0$ 
limit. This approach reveals that the magnetic susceptibility~\eqref{eq:defchi} arises as a smooth limit of susceptibilities with respect to oscillatory
fields. 
As the details of the derivation are somewhat technical, 
we delegate them to App.~\ref{app:derivatives} and only 
quote the main results here.

In the thermodynamic limit, where the momentum variable 
is continuous and the $p_1\to0$ limit can be taken, 
the susceptibility is obtained as
\be
\chi_b =- \lim_{p_1\to0}
\int \dd x_1\, \frac{\cos(p_1x_1)-1}{p_1^2}\, G(x_1) 
= \int \dd x_1\, \frac{x_1^2}{2}\, G(x_1) \,.
\label{eq:chibderiv1}
\ee
In finite volumes ($x_1\in[0,L]$) the momentum variable $p_1$ is discrete, so that the $p_1\to0$ limit does 
not exist. Nevertheless, we can safely employ the formula~\eqref{eq:chibderiv1}
directly in finite volumes, as long as the linear size $L$ is much larger than the 
characteristic length governing the exponential decay of $G(x_1)$. 
Symmetrizing Eq.~\eqref{eq:chibderiv1} to comply with periodic boundary conditions 
and the symmetry $G(x_1) = G(L-x_1)$, 
we arrive at\be
\chi_b = \frac{1}{2} \int_0^L \dd x_1\,  G(x_1)\cdot
\begin{cases}
 x_1^2, &x_1\le L/2 \\
 (x_1-L)^2, & x_1>L/2 
\end{cases}
\,.
\label{eq:Pi0def}
\ee

In the representation~\eqref{eq:Pi0def} the current-current correlator is computed in coordinate space. Only afterwards a Fourier transformation is carried out 
via the convolution with the quadratic kernel in order to represent 
the constant background field. In this way the problem of flux quantization is avoided.
Notice that there is a remnant of flux quantization in the
formula~\eqref{eq:Pi0def}, 
signaled by the cusp in the kernel at $x_1=L/2$. However, this cusp has no practical relevance, 
as in the integral it is multiplied by $G(L/2)$, which is exponentially small.
Thus we do not expect to encounter 
substantial finite volume effects. This is 
contrary to the case of the half-half method~\cite{Levkova:2013qda}, where 
translational invariance is broken already on the level of the 
expectation values, involving a vector potential with kinks.
Nevertheless, we investigate the finite volume effects of the new 
method numerically in Sec.~\ref{sec:res}.

The result~(\ref{eq:chibderiv1}) can be recast into an alternative form 
using the vacuum polarization tensor
\be
\Pi_{\mu\nu}(p) = \int \dd^4 x \,e^{ipx}\expv{j_\mu(x) j_\nu(0)}
\,, \quad\quad
\Pi_{22}(p=\{p_1,0,0,0\}) = -p_1^2\, \Pi(p^2)
\,.
\label{eq:PI22form}
\ee
The second relation, involving the vacuum polarization form factor $\Pi$, only holds for this specific choice of spatial 
indices.\footnote{At zero temperature, the second relation of \protect Eq.~\eqref{eq:PI22form} follows directly from the 
decomposition $\Pi_{\mu\nu}(p)=(p_\mu p_\nu - \delta_{\mu\nu}p^2)\, \Pi(p^2)$. 
For $T>0$ the Lorentz structure of $\Pi_{\mu\nu}(p)$ is more complicated
so that, in addition to $\Pi$, a 
form factor $\Pi_L$ appears~\cite{bellac2000thermal,kapusta2006finite}.
However, in the static case ($p_4=0$) only $\Pi$ contributes to
the spatial components
of $\Pi_{\mu\nu}$ so that the second relation of Eq.~\eqref{eq:PI22form}
continues to hold.}
Employing these definitions, we can rewrite 
\be
\chi_b = \lim_{p_1\to0}\,
\Pi(p^2) =\Pi(0),\quad\quad
\Pi(p^2) = 
\int \dd x_1\, \frac{1-\cos(p_1x_1)}{p_1^2}\, G(x_1) \,,
\label{eq:chiPi}
\ee
where we used that the imaginary part of $\Pi(p^2)$ vanishes. 

We note that the vacuum polarization function
$\Pi$ has been the subject of intense research as it appears in 
the hadronic contribution to the muon anomalous magnetic moment, see, e.g., the recent review~\cite{Meyer:2018til}. In that setting the relevant observable is 
the second moment of the two-point
function of the electromagnetic
current~\eqref{eq:current}, projected to zero spatial momentum~\cite{Francis:2013fzp}. Exchanging
the time coordinate $x_4$ for the spatial coordinate $x_1$, one can
obtain $\chi_b$ in an analogous way in our background field setup~\cite{Bali:2015msa}.
The two determinations are equivalent at zero temperature. For $T>0$
it is important to use spatial momenta, i.e.\ a kernel involving 
spatial coordinates for $\chi_b$, since this encodes the magnetic 
response.

In App.~\ref{app:derivatives} we derive a similar representation for $\tau_{fb}$ as well.
In this case the equivalents of Eqs.~\eqref{eq:G22def} and~\eqref{eq:Pi0def} become\footnote{On general grounds, the linear response of the
    expectation value of
  an $n$-point function with respect to a background field can always
  be obtained by computing $(n+1)$-point functions in the vacuum. Usually, the
  former method is favorable in terms of the statistical noise.
  However, the latter option exempts us from the need of generating additional
  gauge ensembles with non-vanishing values of the background field. As already
  discussed, in the present context, we also circumvent the issue
  of flux quantization.
}
\be
\tau_{fb} = \frac{i}{q_f/e}\int_0^L\! \dd x_1\,  H_f(x_1) \cdot
\begin{cases}
 x_1, &x_1\le L/2 \\
 x_1-L, & x_1>L/2 
\end{cases}\,, \quad
H_{f}(x_1) = \int\! \dd x_2 \,\dd x_3 \,\dd x_4 \,\expv{\bar\psi_f\sigma_\subs\psi_f(x)j_2(0)} \,.
\label{eq:Hdef_tau}
\ee
We remark that the second moment of the photon
DA is accessible too, replacing $\sigma_\subs$ by
combinations of $\sigma_{\mu\nu}\left(\overleftarrow{D}_{\rho}\overleftarrow{D}_{\sigma}+\overrightarrow{D}_{\rho}\overrightarrow{D}_{\sigma}-2
\overleftarrow{D}_{\rho}\overrightarrow{D}_{\sigma}\right)$ that are
antisymmetric in indices equal to $1$ and $2$, symmetrized
over all other non-trivial combinations of indices and with all
traces subtracted, see, e.g., Ref.~\cite{Braun:2016wnx}.
This is beyond the scope
of the present work.

In summary, via the relation~\eqref{eq:Pi0def} we are able to determine the magnetic susceptibility using direct measurements at $B=0$. This is certainly advantageous over 
calculating the free energy density (which cannot be obtained as a simple 
expectation value) at nonzero magnetic fields and differentiating it numerically.
The similar relation for the tensor coefficient, Eq.~\eqref{eq:Hdef_tau}, might also be used 
to avoid measuring $\expv{\bar\psi_f\sigma_\subs\psi_f}$ at $B>0$. However, since the latter 
is a simple one-point function, the gain is not obvious in this case. 
In App.~\ref{app:derivatives} we compare the two methods 
for this observable and conclude that the 
correlator method indeed gives larger statistical errors. Therefore, we opted to 
use our earlier 
results~\cite{Bali:2012jv} for $\expv{\bar\psi_f\sigma_\subs\psi_f}$.

\section{Results}
\label{sec:res}

\begin{figure}[t]
 \centering
 \mbox{
 \includegraphics[width=8.5cm]{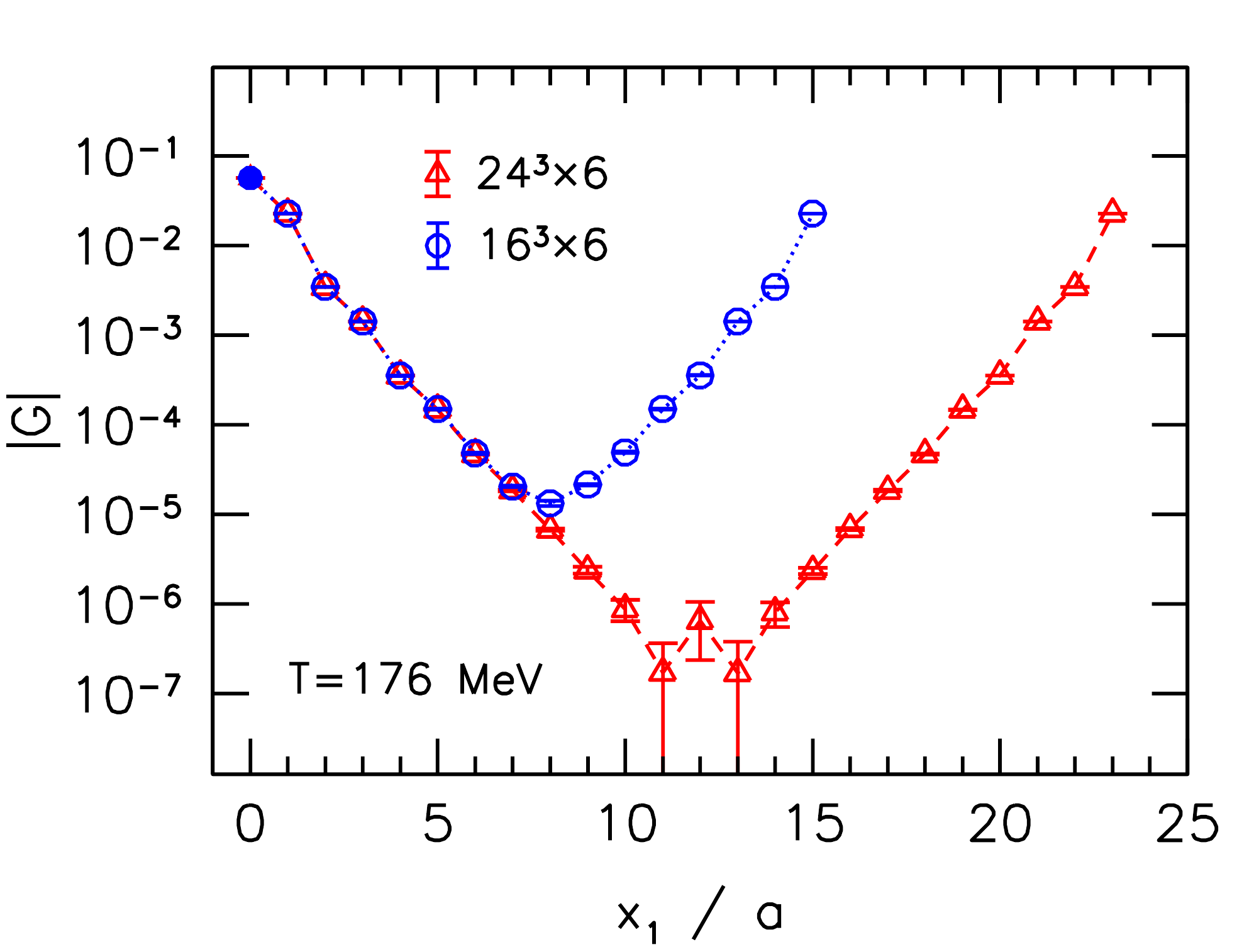}
 \includegraphics[width=8.5cm]{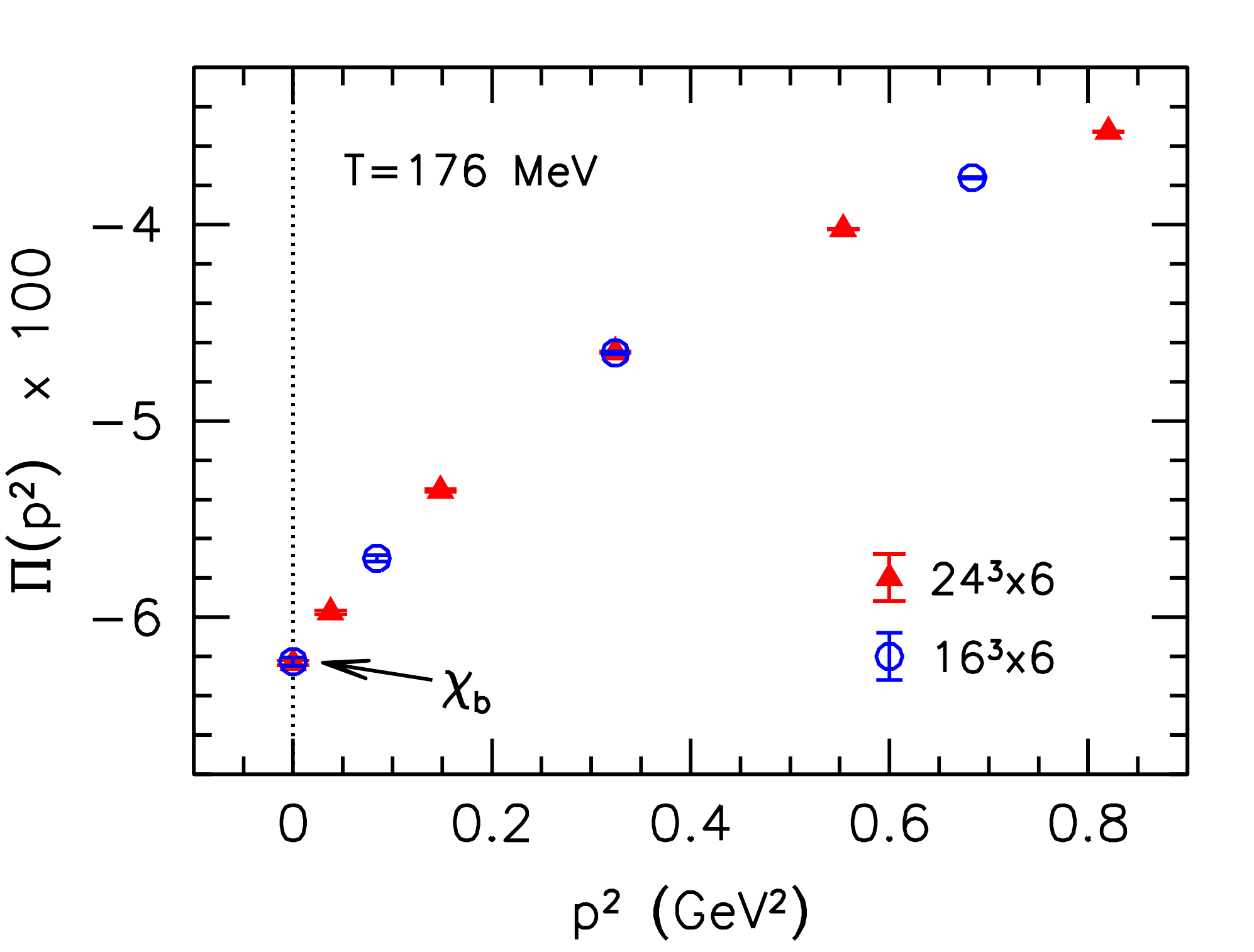} }
 \caption{\label{fig:voldep}Left panel: comparison of the absolute 
 value of the current-current correlator for two different volumes, $24^3\times 6$ (red) and $16^3\times 6$ (blue). Filled (open) points indicate positive (negative) values.
 Right panel: the vacuum polarization function at spatial momenta
 for $T\approx 176 \textmd{ MeV}$ using two different volumes.
 The bare magnetic susceptibility can be read off from the intersect $\Pi(0)$.
}
\end{figure}

First we demonstrate that -- in accordance with Eq.~\eqref{eq:chiPi} --
$\chi_b=\Pi(0)$ arises as a smooth 
limit of the vacuum polarization function $\Pi(p^2)$ at 
spatial momenta. 
To this end we calculate 
the correlator $G(x_1)$ using $\mathcal{O}(1000)$ random sources located on three-dimensional $x_1$-slices 
of our lattices, taking into account both connected and disconnected 
contributions. The correlator is shown in the left panel of Fig.~\ref{fig:voldep} for our $N_t=6$ lattices at a high temperature $T\approx 176\textmd{ MeV}$. Here we compare two different volumes with $N_s=24$ and $N_s=16$, revealing that finite size effects in the 
exponential fall-off are tiny. 
Subsequently, $G(x_1)$ is convoluted 
with the kernels of Eqs.~\eqref{eq:Pi0def} and~\eqref{eq:chiPi} 
to obtain $\Pi(0)$ and $\Pi(p^2)$, respectively. 
In the right panel of Fig.~\ref{fig:voldep} we show $\Pi(p^2)$ for low momenta, again at the same temperature $T\approx 176 \textmd{ MeV}$. 
As expected, the zero-momentum limit is approached smoothly 
and the two volumes are 
found to agree perfectly.

\begin{figure}[t]
 \centering
 \mbox{
 \includegraphics[width=8.5cm]{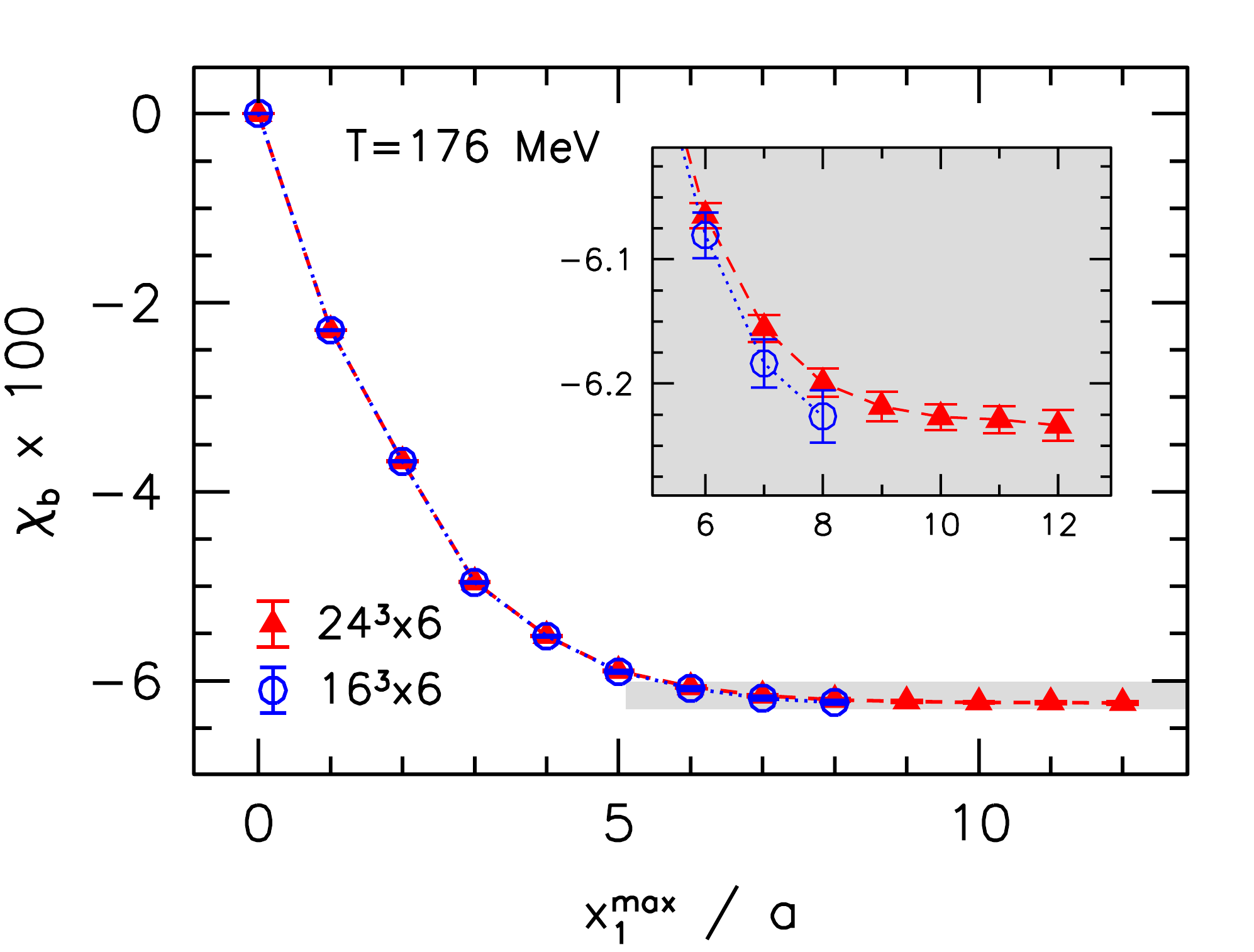}
 \includegraphics[width=8.5cm]{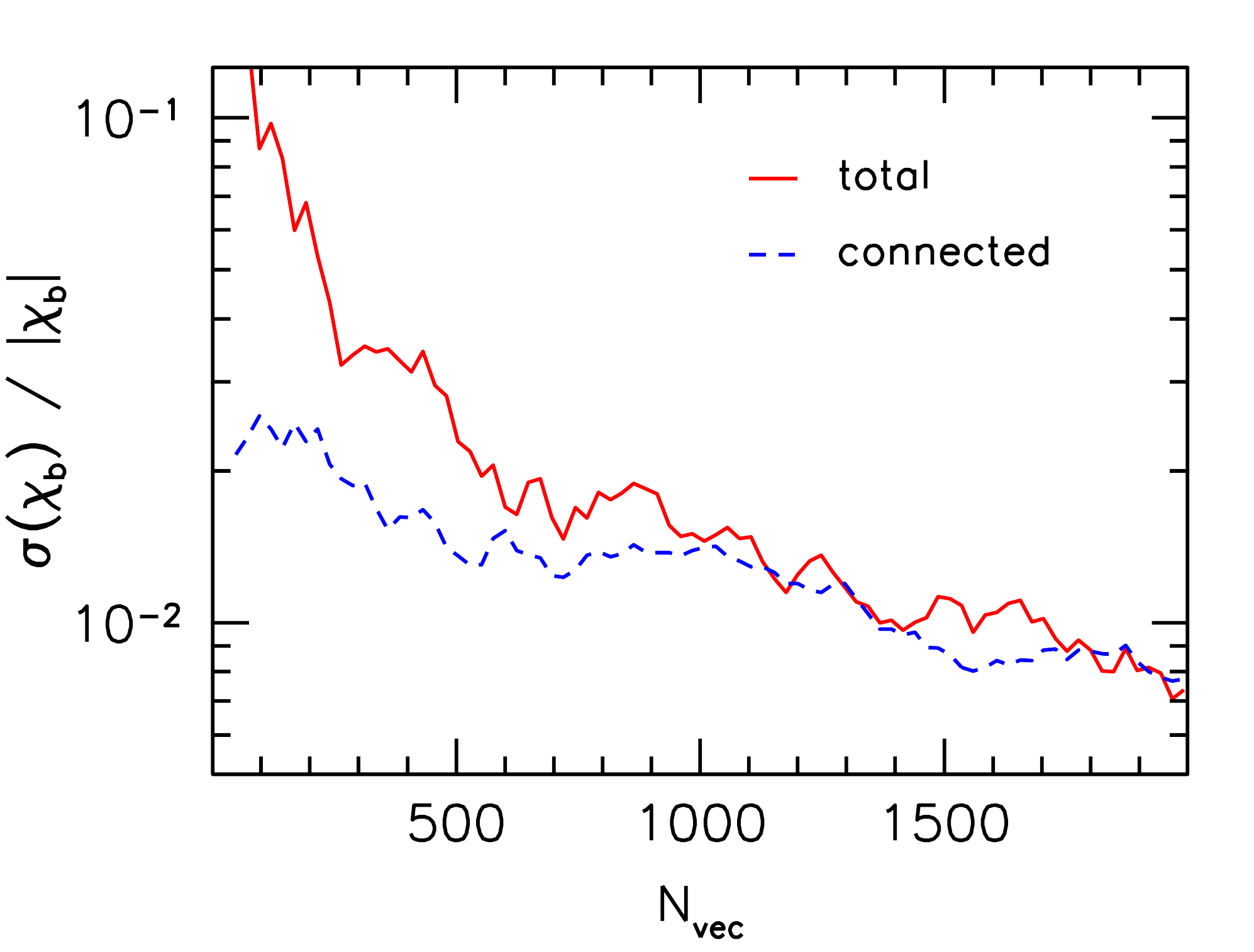} }
 \caption{\label{fig:voldep_contr}Left panel: the susceptibility obtained via a truncation of \protect Eq.~\eqref{eq:Pi0def} for two different volumes, $24^3\times 6$ (red) and $16^3\times 6$ (blue). 
 The inset zooms into the region near $x_1^{\rm max}=N_sa/2$. 
 Right panel: relative error of the susceptibility as a function 
 of the number of employed noisy estimators.
}
\end{figure}

Next we investigate finite volume effects in more detail. 
In particular, we truncate the convolution~\eqref{eq:Pi0def} 
at $x_1^{\rm max}\le L/2$ and plot the so-obtained truncated 
susceptibility in the left panel of Fig.~\ref{fig:voldep_contr}.
This sheds more light on why volume effects are so small.
While for the smaller volume, 
the exponential decay is cut off at a lower $x_1$, the slight 
enhancement of $G$ around $L/2$ due to the backward propagating exponential (see the left panel of Fig.~\ref{fig:voldep}) almost completely corrects for this.
Finally, we estimated the deviation of the result from the 
thermodynamic limit by considering a single-exponential fit 
of the correlator at $x_1<L/2$ and performing the convolution~\eqref{eq:chibderiv1} for $L/2\le x_1<\infty$. 
For $N_s=16$ this correction is found to be about half of the statistical error of $\chi_b$, while
for $N_s=24$ it is found to be two orders of magnitude smaller
than that. 
For these analyses we 
considered the results at $T\approx 176\textmd{ MeV}$, where 
we have very precise data. For the lower temperatures our data 
are noisier; here we find
the finite volume errors to be significantly smaller than 
our statistical uncertainties already for $N_s=16$.

Before turning to the main results, we discuss the statistical accuracy 
and the numerical costs
of the present method. 
The right panel of Fig.~\ref{fig:voldep_contr} shows the relative error of $\chi_b$ 
at $T\approx 113 \textmd{ MeV}$ on our $24^3\times 6$ lattices 
as a function of the employed number $N_{\rm vec}$ of noisy 
estimators. The evaluation of $G(x_1)$ requires two inversions 
for the light quarks and two for the strange quark for each 
noisy estimator. 
The figure reveals that sub-percent errors can be achieved.
For sufficiently high $N_{\rm vec}$ the connected contributions are found to dominate the error, as already 
recognized in Ref.~\cite{Bali:2015msa}.
Alternative methods to calculate $\chi_b$~\cite{Bonati:2013lca,Bonati:2013vba,Bali:2014kia} require several independent simulations at nonzero $B$ and a reconstruction of $\log\Z$ for each magnetic field 
and are therefore much more expensive than the present
approach (of course, in that case the physical $B>0$ ensembles might also 
be used for other purposes). To be specific, we consider our results~\cite{Bali:2014kia} using 
the integral method at the same temperature and lattice spacing 
as above. In that case we needed to perform
around 40 independent simulations (at nonzero $B$ as well as at different 
quark masses), generating several hundred 
decorrelated configurations and measuring the quark condensate on 
each ensemble. We achieved a relative error of about four percent.
More importantly, the present method outperforms previous 
alternatives because this determination of $\chi_b$ entails 
no further systematic uncertainty, unlike
approaches~\cite{Bonati:2013lca,Bonati:2013vba,Bali:2014kia}, where a numerical differentiation of $\log\Z(B)$ is required.

\subsection{The magnetic susceptibility}
\label{sec:res_1}

\begin{figure}[b]
 \centering
 \includegraphics[width=8.5cm]{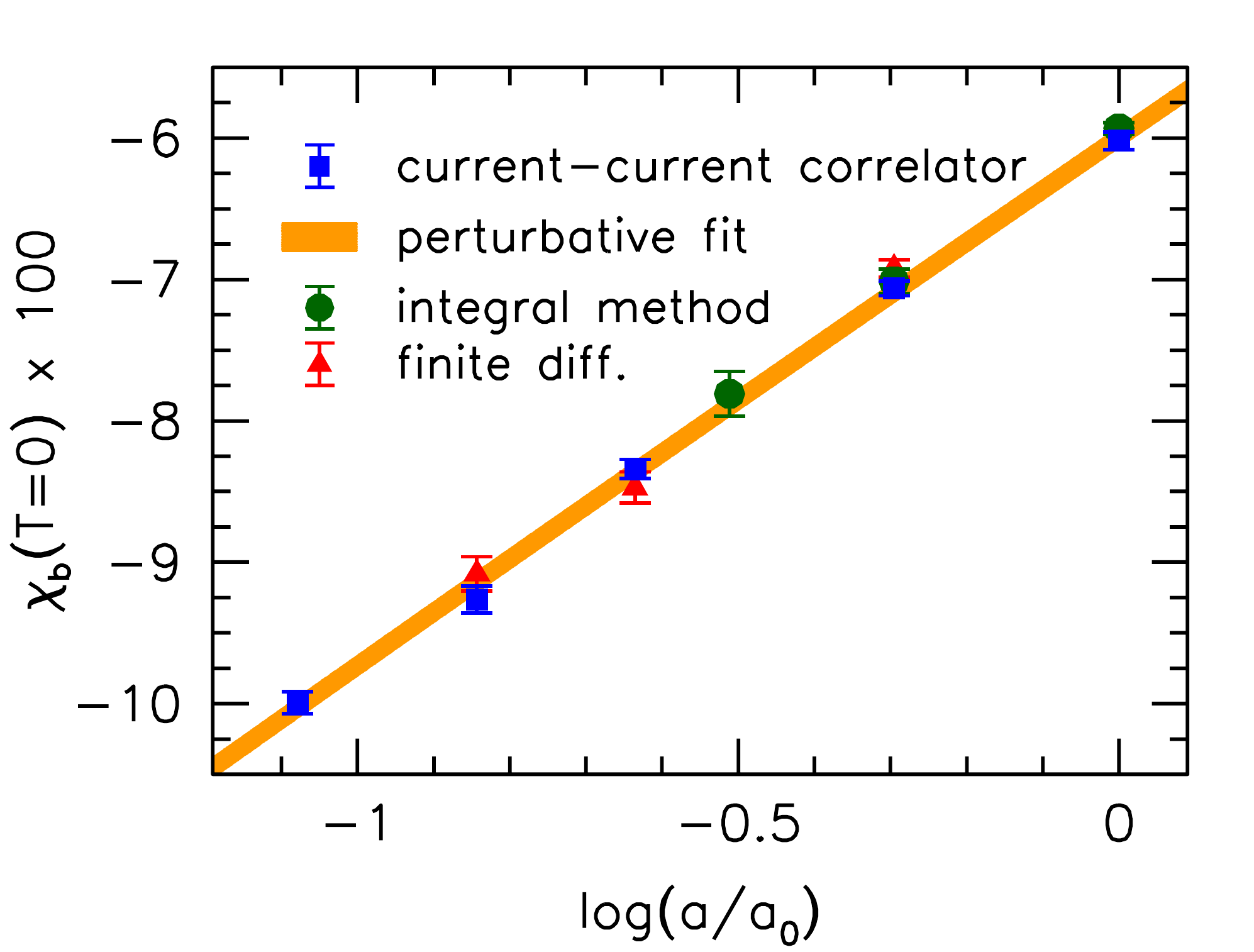}
 \caption{\label{fig:chi0}Bare magnetic susceptibility at zero temperature 
 versus the logarithm of the lattice spacing, normalized to $a_0=1.46\textmd{ GeV}^{-1}$. Different 
 approaches are compared: the finite difference method~\protect\cite{Bonati:2013vba} (red triangles), the generalized integral method~\protect\cite{Bali:2014kia} (green circles)
 and the new approach via current-current correlators (blue squares). The 
 orange band indicates the fit based on perturbation theory, Eq.~\protect\eqref{eq:formula}.}
\end{figure}
  
We compare the results of our new method for $\chi_b$ to our old data (generalized integral method)
and also to those of Ref.~\cite{Bonati:2013vba} (finite difference method)
in  Fig.~\ref{fig:chi0} at zero temperature.\footnote{
  Ref.~\cite{Bonati:2013vba} employs the same lattice action.
  For a different action the bare susceptibilities would not only differ
  in terms of lattice artifacts but also by an additive constant, due 
  to the different choice of renormalization scheme.
}
Within errors perfect agreement between the three groups of results is found. 
The data -- plotted in Fig.~\ref{fig:chi0} against $\log(a)$ -- clearly reflect the logarithmic divergence dictated by Eq.~\eqref{eq:b1}. 
Similarly to our fitting strategy in Ref.~\cite{Bali:2015msa}, 
here we also include the universal perturbative 
QCD corrections to the QED $\beta$-function coefficient
$c_1$~\cite{Baikov:2012zm}, see Eq.~\eqref{eq:beta1def},
  where $g^2=6/\beta$ is obtained from the inverse lattice
  coupling $\beta$ at the lattice scale $a^{-1}$.
We also take into account $\mathcal{O}(a^2)$ lattice artifacts so that our
fit function reads
\be
\chi_b = 2\beta_1(a^{-1}) \cdot \big[ \log(a/a_0) + \log(\muqed a_0) \big] \cdot \left[1+z_1 (a/a_0)^2\right], \quad\quad a_0=1.46 \textmd{ GeV}^{-1}.
\label{eq:formula}
\ee
The result of this fit, with the parameter values
\be
\muqed = 115(3)(5) \textmd{ MeV}\,, \quad\quad z_1=-0.05(1),
\label{eq:mures}
\ee
is shown as an error band in Fig.~\ref{fig:chi0}.
We also considered fits with further (quartic) lattice artifacts. 
The impact of this is included in the second parentheses
of Eq.~\eqref{eq:mures} 
for $\muqed$ as a systematic error.
The renormalization scale agrees within errors with our earlier determinations~\cite{Bali:2014kia,Bali:2015msa}.
It also lies near the mass of the lightest charged hadron (the 
charged pion) which effectively sets the scale for the magnetic 
response of this system. Nevertheless,
note that the value of $\muqed$ depends on the choice of the regulator.

\begin{figure}[t]
 \centering
 \mbox{
 \includegraphics[width=8.5cm]{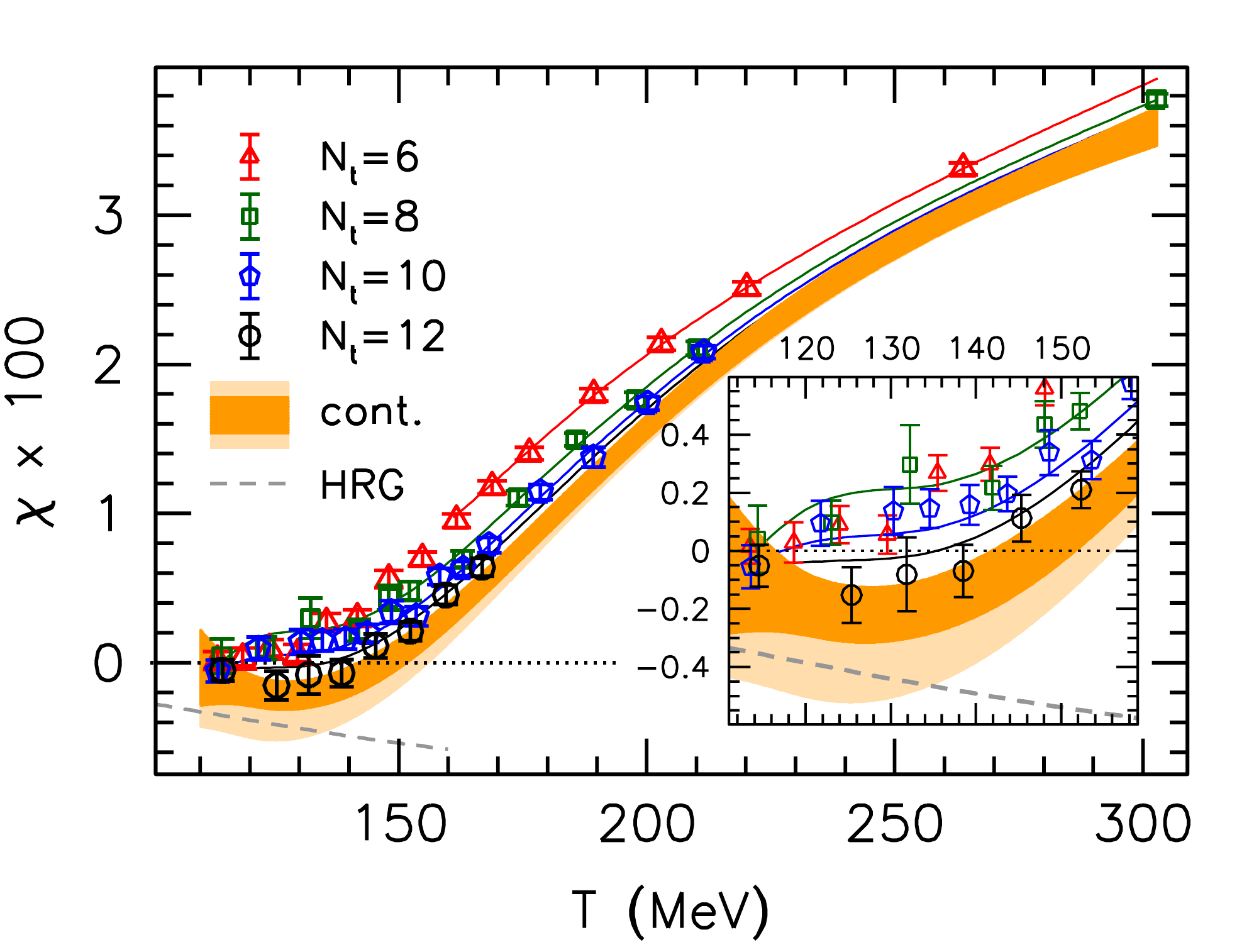}
 \includegraphics[width=8.5cm]{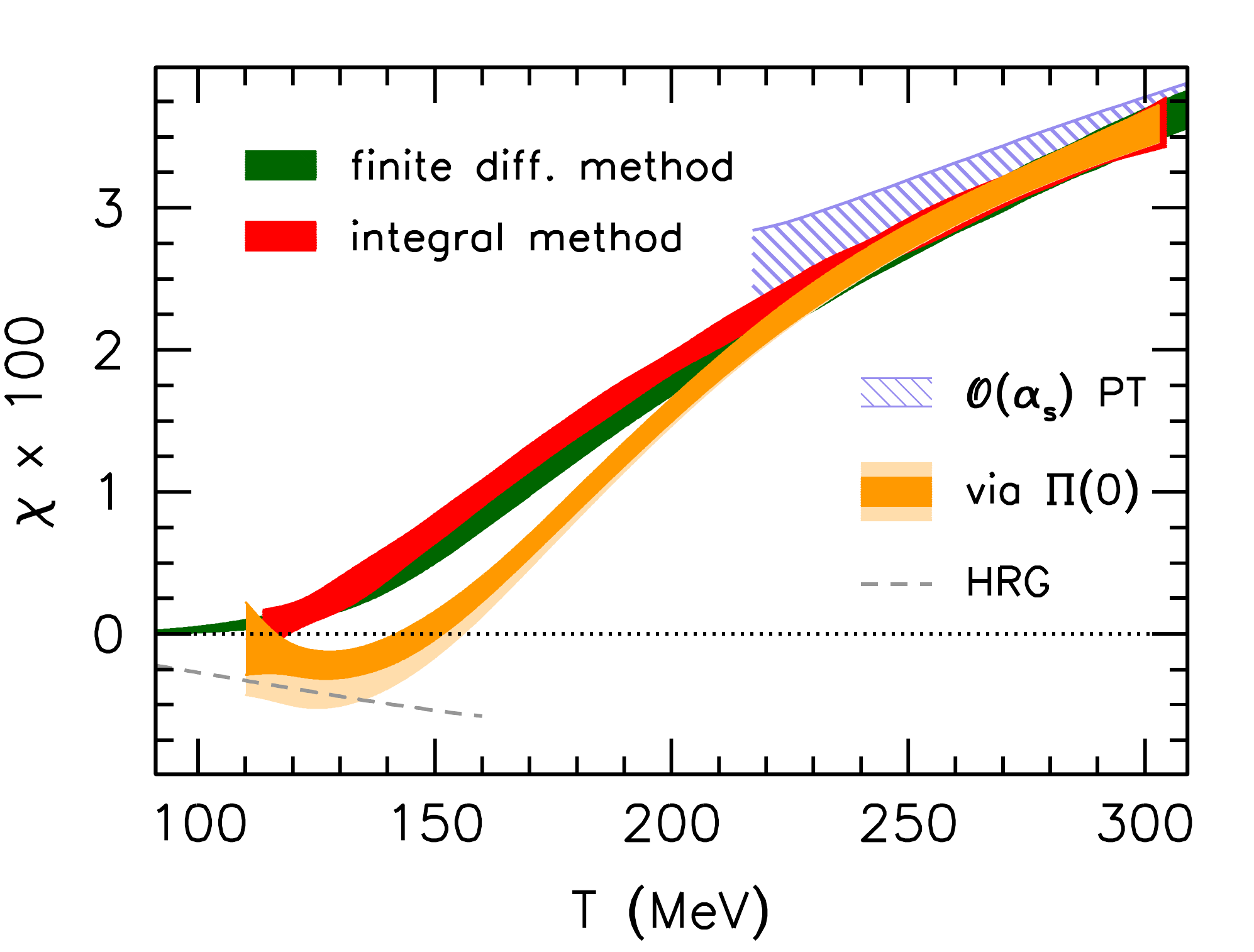}
 }
 \caption{\label{fig:chiT}Left panel: renormalized susceptibility at nonzero temperature. The symbols indicate different lattice spacings and 
 the dark orange band the continuum limit. The light orange 
 band represents an estimate of systematic errors of the continuum 
 extrapolation.
 The dashed gray line is the HRG model prediction~\protect\cite{Bali:2014kia}. The inset zooms into the low-temperature region
 to highlight the diamagnetic response there.
 Right panel: our results for $\chi$ (orange, labeled ``via $\Pi(0)$'') are 
 compared to the results of Ref.~\protect\cite{Bonati:2013vba} (green)
 and those of Ref.~\protect\cite{Bali:2014kia} (red) as well as
 to the HRG model prediction (dashed gray line)
 and to perturbation theory~\protect\cite{Bali:2014kia} (dashed light blue band),
 see Eqs.~\protect\eqref{eq:highTchi} and~\protect\eqref{eq:beta1def}.}
\end{figure}

The formula~\eqref{eq:formula} is used to interpolate $\chi_b$
and then employed to 
renormalize the susceptibility at nonzero temperatures 
according to Eq.~\eqref{eq:renorm}: $\chi=\chi_b(T)-\chi_b(0)$.
The results are shown in the left panel of Fig.~\ref{fig:chiT}
for a broad range of temperatures and 
four lattice spacings $a=1/(N_tT)$ with $N_t=6,8,10$ and $12$.
A continuum extrapolation is performed by means 
of a multi-spline fit~\cite{Endrodi:2010ai} 
taking into account $\mathcal{O}(a^2)$ lattice 
artifacts. To have acceptable fits of this type, 
we needed to discard 
the coarsest lattices ($N_t=6$ points at
$T\lesssim 160\textmd{ MeV}$). 
We also repeated the analysis including $\mathcal{O}(a^4)$ discretization 
errors as well, this time fitting all available data points. 
The systematic error was estimated by the difference of 
these two extrapolations as well as by varying the spline node
points and by including/excluding $N_t=6$ data points at high temperatures for the $\mathcal{O}(a^2)$ fit.
In addition, we consider a further systematic error due to
lattice artifacts 
related to the taste splitting of the staggered spectrum.
This effect is particularly relevant at low temperatures and can be estimated 
by a generalization of the Hadron Resonance Gas (HRG) model that we 
describe in App.~\ref{sec:B}. The light yellow bands in both panels
of Fig.~\ref{fig:chiT} indicate the total systematic uncertainties.

The results demonstrate strong paramagnetism in the quark-gluon
  plasma
phase, in agreement with previous lattice studies~\cite{Bonati:2013vba,Bonati:2013lca,Bali:2013owa,Bali:2013esa,Levkova:2013qda,Bali:2014kia}.
At high temperatures the results are well described by the free theory, which predicts (see App.~\ref{app:freecase})
\be
\chi(T) = \beta_1(\mu_{\rm therm}) \cdot \log \left(\gamma\,\frac{T^2}{\muqed^2}\right) + \mathcal{O}(1/T^2)\,,
\label{eq:highTchi}
\ee
where
$\gamma=\mathcal{O}(1)$ is a regulator-dependent
constant.\footnote{For example in the free case with cut-off
  regularization $\gamma=\pi^2 e^{-\gamma_E}$, see App.~\ref{sec:highTexp}.
  The renormalized susceptibility $\chi(T)$ and the ratio
    $\gamma/\muqed^2$ are regulator-independent.
}
As indicated, QCD corrections at the thermal scale 
$\mu_{\rm therm}$ affect the leading behavior.
In the right panel of Fig.~\ref{fig:chiT}
we also include a comparison to this perturbation
  theory formula,
with the scheme-independent
$\mathcal{O}(\alpha_s)$ corrections to $\beta_1$ taken into
account~\cite{Bali:2014kia}. 
Here we
set $\gamma=1$ and
use the $\overline{\mathrm{MS}}$ scheme definition of the
  strong coupling $\alpha_s=g^2/(4\pi)$, running this at
  five-loop order~\cite{Baikov:2016tgj} to the thermal scale
  $\mu_{\rm therm}\sim 2\pi T$. We employ
  the central value $\Lambda_{\rm QCD}^{\overline{\rm MS}}=0.341\textmd{ GeV}$
  from the recent three flavor QCD determination of $\alpha_s$ by the ALPHA
  Collaboration~\cite{Bruno:2017gxd}.
  The band in the figure corresponds to a variation of the thermal scale
  from $\pi T$ to $4\pi T$. The perturbative formula 
agrees surprisingly well with our results down to
temperatures $T\sim 200\textmd{ MeV}$.

\begin{figure}[t]
 \centering
 \mbox{
 \includegraphics[width=8.5cm]{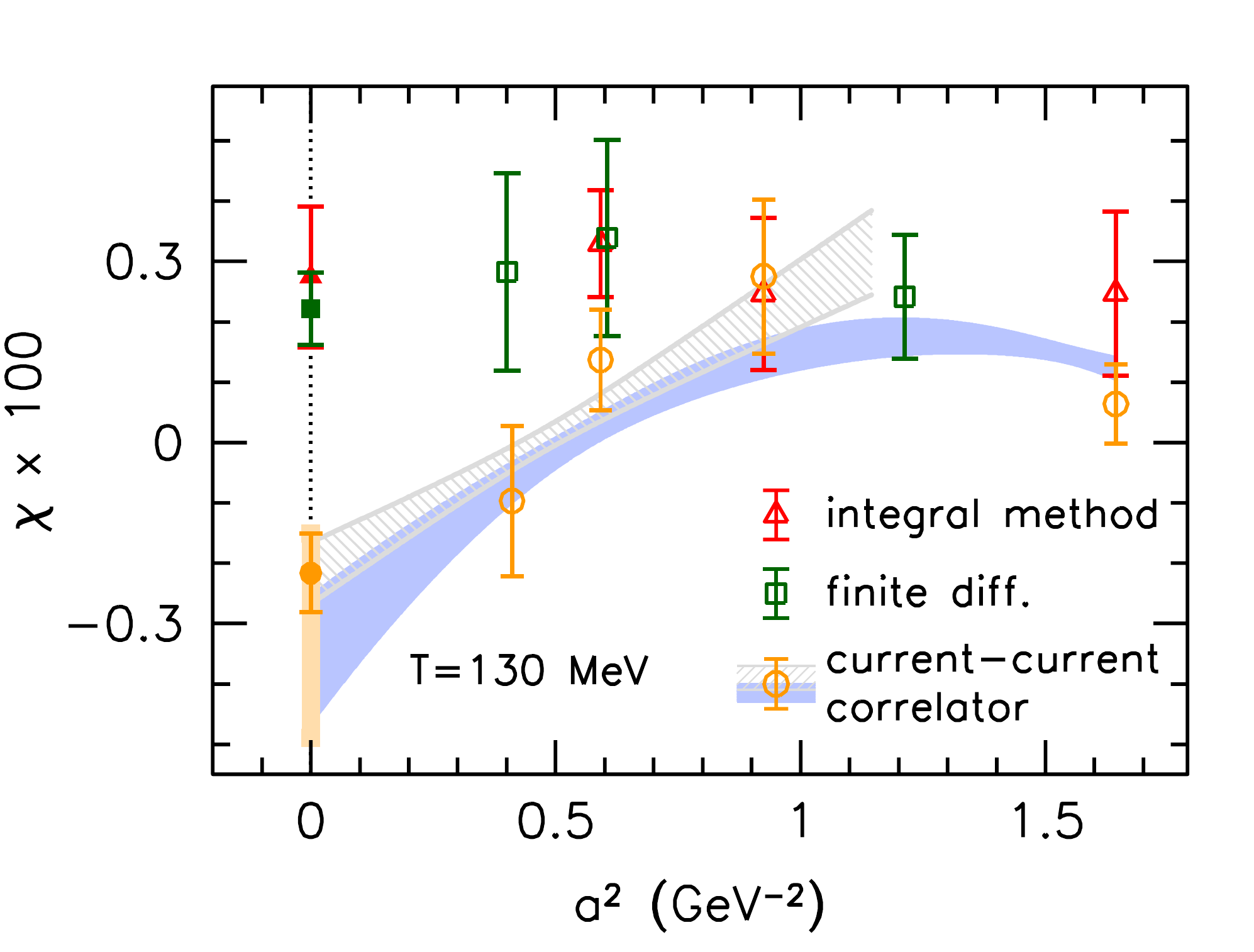}
 \includegraphics[width=8.3cm]{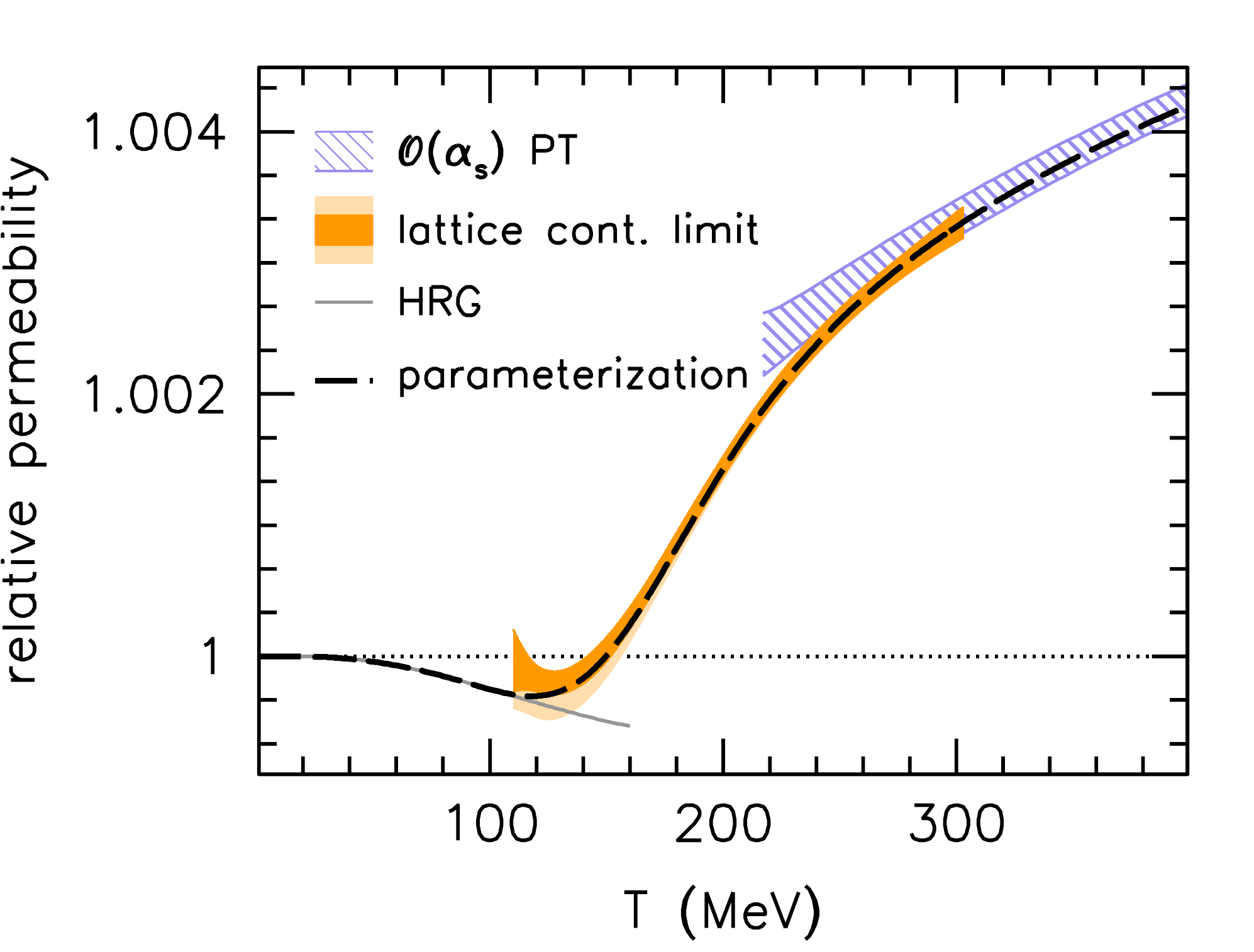}
 }
 \caption{\label{fig:artefacts}
 Left panel: lattice discretization errors in the renormalized magnetic susceptibility. 
 The results of Ref.~\protect\cite{Bonati:2013vba} (green) and those of Ref.~\protect\cite{Bali:2014kia} (red)
 are compared to the present approach (yellow) including systematic uncertainties (light yellow). The dashed gray and the light blue 
 bands represent the $T=130\textmd{ MeV}$ slices of the multi-spline fit involving 
 up to $\mathcal{O}(a^2)$ and $\mathcal{O}(a^4)$ lattice artefacts, 
 respectively.
 Note that for the green points at $a>0$, 
 $\chi$ was obtained by 
 temperature-interpolations of the 
 results published in Ref.~\protect\cite{Bonati:2013vba}.
 Right panel: parametrization of the relative magnetic permeability
 $\mu/\mu_0=(1-e^2\chi)^{-1}$
 via the function~\protect\eqref{eq:chipars} of App.~\protect\ref{app:parameterization}.
 }
\end{figure}

In contrast to the paramagnetic behavior at high $T$, towards low temperatures
the continuum extrapolated results become negative,
revealing a diamagnetic response, previously noted in Ref.~\cite{Bali:2014kia}. This behavior is in agreement, albeit within large errors, with 
the HRG model prediction~\cite{Bali:2014kia} (dashed line in the figures). 
In the right panel of Fig.~\ref{fig:chiT} we also include
results obtained from other approaches that employed the
same lattice action.
Around the pseudo-critical temperature $T_c\approx 155 \textmd{ MeV}$
a significant 
difference is visible between earlier determinations and our present results.
Ref.~\cite{Bonati:2013vba} carried out a continuum 
extrapolation using fixed $\beta$ ensembles with lattice spacings 
$a\ge 0.125 \textmd{ fm}$. In Ref.~\cite{Bali:2014kia} we used the fixed $N_t$ approach with
$N_t=6,8,10$ ensembles, while 
in the present study $N_t=6,8,10,12$ lattices are simulated.
To highlight the differences between the continuum extrapolations, in
the left panel of Fig.~\ref{fig:artefacts} we 
plot the lattice spacing-dependence of the susceptibility for all three approaches.
We pick one temperature $T=130\textmd{ MeV}$, where the deviation of the continuum estimates 
is substantial.

The left panel of Fig.~\ref{fig:artefacts} reveals the importance of our new $N_t=12$ ensemble, 
showing a significant downward trend as the lattice spacing is reduced and 
a negative value in the continuum limit. The downward trend is not captured by 
our previous estimate using the integral method on $N_t\le 10$ lattices~\cite{Bali:2014kia}, 
neither is it visible in the data of Ref.~\cite{Bonati:2013vba}. 
We note that in the left panel of Fig.~\ref{fig:artefacts}, a difference beyond one standard deviation can only be observed at the 
smallest lattice spacing. Nevertheless, our $N_t=12$ data lie 
consistently below the other lattice spacings for all temperatures 
(see the left panel of Fig.~\ref{fig:chiT}) so that the downward trend towards $a\to0$ 
is statistically significant. We indeed expect lattice artefacts in $\chi$
to be large and positive in this temperature region, as predicted 
by the generalized HRG model of App.~\ref{sec:B}.
Finally we remark that Ref.~\cite{Bonati:2013vba} performed the continuum extrapolation
assuming a strictly positive function for $\chi(T)$. Excluding the possibility 
of a negative susceptibility in the continuum limit might in general  underestimate the systematics of the 
extrapolation. 
To clarify this issue, dedicated simulations
should be performed with the same action using all available 
methods,
preferably at the same temperatures and the same values of the lattice spacing.

Finally we provide a parametrization that connects all three approaches 
(HRG, lattice continuum limit and perturbation theory) and 
describes $\chi$ for arbitrary temperatures. The details 
are discussed in App.~\ref{app:parameterization}.
In the right panel of Fig.~\ref{fig:artefacts} we plot this parametrization, 
translated to the magnetic permeability $\mu/\mu_0=(1-e^2\chi)^{-1}$,
expressed in units of the vacuum permeability $\mu_0$.
This combination is equal to the ratio of the 
magnetic induction and the external field, see,
e.g., Refs.~\cite{Bonati:2013lca,Bali:2014kia}.

\subsection{The normalization of the photon distribution amplitude}
\label{sec:res_2}
Here we address the tensor coefficients $\tau_{fb}$ at zero temperature.
We consider a set of independent gauge ensembles, generated at the physical 
value of the strange quark mass $m_s=m_s^{\rm phys}$, but at different values
of the light quark mass: $0.5 \,m_{ud}^{\rm phys} \le m_{ud} \le m_s^{\rm phys}$.
We follow a similar strategy as
in Ref.~\cite{Bali:2012jv}, simultaneously
fitting the dependence of $\tau_{ub}\cdot Z_T$ on the light quark 
mass $m_{ud}$ and on the lattice spacing $a$ according to the ansatz~\eqref{eq:t1}.
Since $Z_T$ is found to depend very mildly on the lattice spacing within
the range covered (see 
Fig.~\ref{fig:renconsts} of App.~\ref{sec:renormconstants}), 
this does not significantly affect the functional dependence on $a$.
We note that on our coarsest ensembles the uncertainty of $Z_T$ is 
quite large, which also imprints on the errors of the renormalized
tensor coefficients.

\begin{figure}[th]
 \centering
 \includegraphics[width=8.5cm]{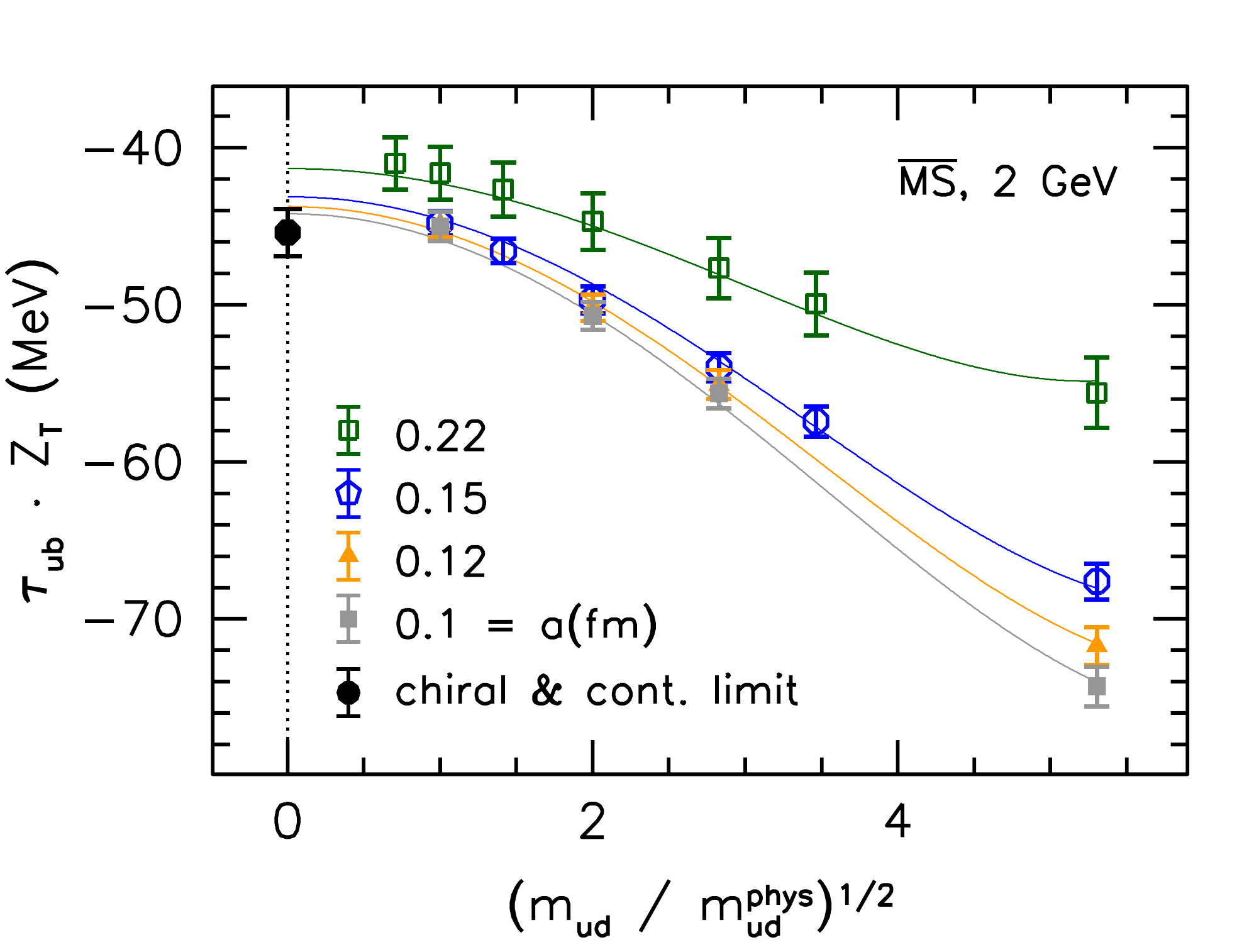}
 \caption{\label{fig:tauuT0}Light quark mass-dependence of the tensor coefficient at $T=0$
   for the up quark using our four finest lattice spacings (green to gray symbols). The index $b$ indicates that the QED divergence that one encounters
   at $m_{ud}>0$ has not been subtracted. The results diverge logarithmically
   towards the continuum 
   limit for any $m_{ud}\neq0$. In contrast, the chiral limit 
   is free of ultraviolet divergences and a combined chiral and 
   continuum limit exists (black circle).}
\end{figure}

Notice that $\tau_{fb}$ diverges for $a\to0$ for any quark mass, 
except in the chiral limit, where it is ultraviolet-finite. 
This tendency is clearly visible in Fig.~\ref{fig:tauuT0}, which 
shows our results for the up quark.
Therefore, we can define an ultraviolet-finite observable
for the light quarks, without any zero-temperature subtraction,
namely the chiral limit of the 
tensor coefficient. In contrast, to calculate $\chi^{\rm spin}$ we will 
need to take differences between results obtained at
different temperatures (see below).

The ansatz~\eqref{eq:t1} contains the free parameters $\tau_f$ and $\muqed$.
In addition, we include a quadratic mass-dependence and 
lattice artifacts of $\mathcal{O}(a^2)$ to each parameter in the fit.
Varying the fit ranges in $a$ and in $m_{ud}/m_{ud}^{\rm phys}$ as well as the 
functional form, we carry out several acceptable fits that are used to build
a histogram 
for the chiral continuum limit of the tensor coefficient.
In this combined limit we obtain in the $\overline{\mathrm{MS}}$ scheme
\be
T=0:\quad\quad
f_{\gamma}^{\perp}(2\textmd{ GeV})\equiv
\lim_{m_{ud}\to0}\tau_{u b} \cdot Z_T(2\textmd{ GeV})  =-45.4(1.5) \textmd{ MeV}\,.
\ee
The central value differs from our previous result~\cite{Bali:2012jv}
  $f_{\gamma}^{\perp}=-40.3(1.4) \textmd{ MeV}$, mainly
  due to the multiplicative renormalization factor that we
  determined non-perturbatively here, see Fig.~\ref{fig:renconsts}
  in App.~\ref{sec:renormconstants}.

\begin{figure}[t]
 \centering
\mbox{
 \includegraphics[width=8.5cm]{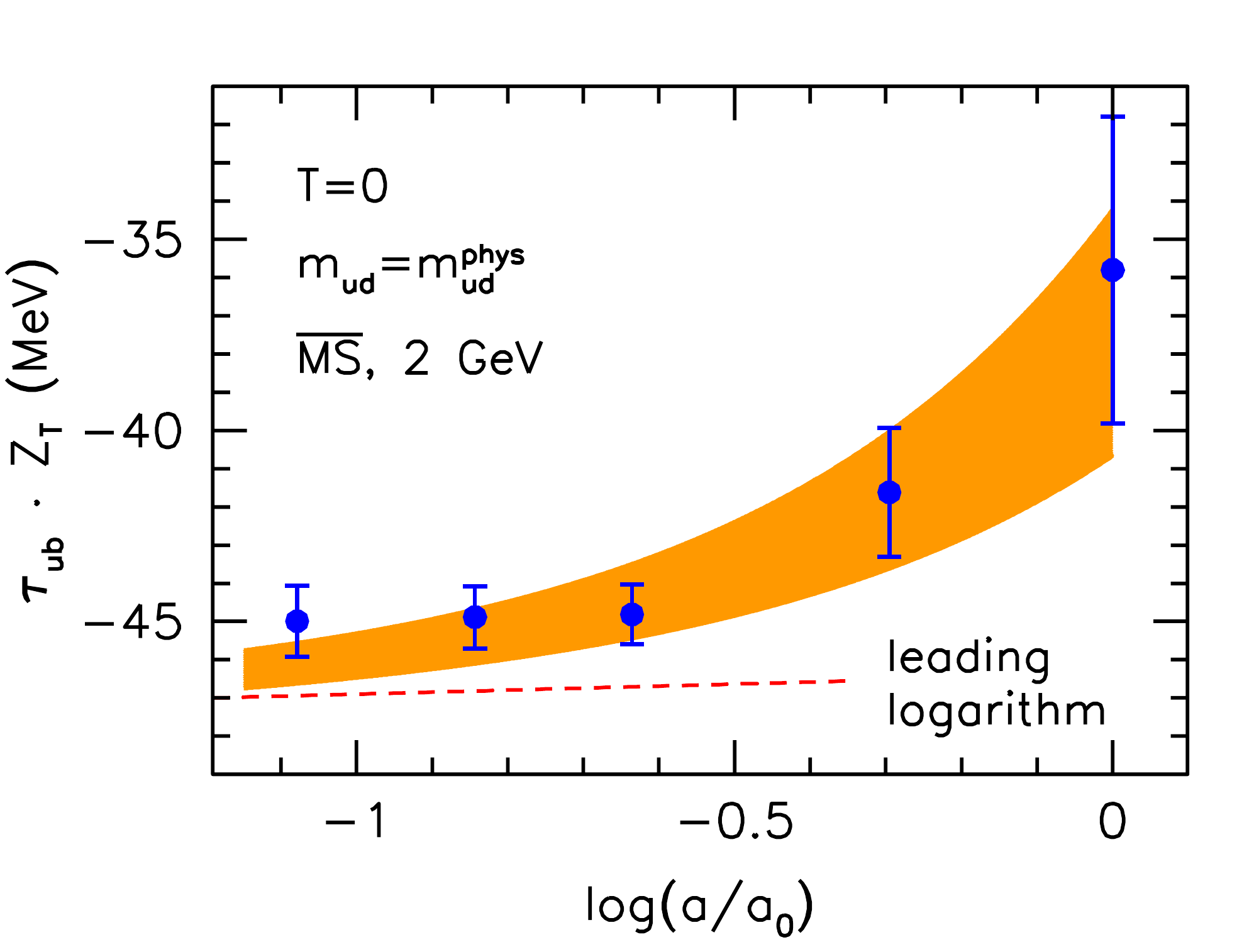}
 \includegraphics[width=8.5cm]{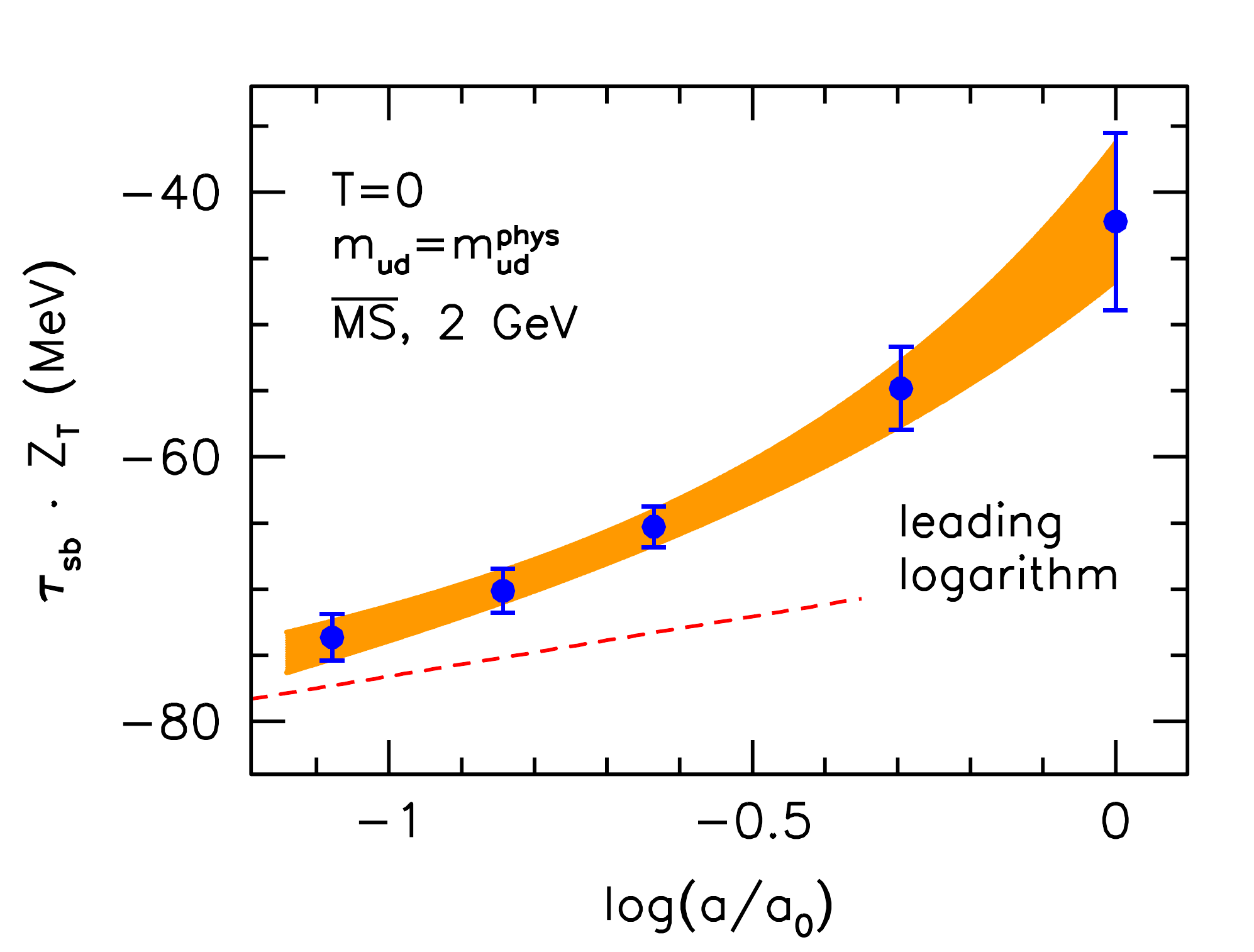}
 }
 \caption{\label{fig:mphys_adep}The bare tensor coefficient for the up quark (left panel) and 
 for the strange quark (right panel) at the physical point and at zero temperature (blue points), together with an interpolation (orange bands).
 For the up quark this interpolation is the $m_{ud}=m_{ud}^{\rm phys}$ 
 slice of a two-dimensional fit like in Fig.~\protect\ref{fig:tauuT0}.
 The red dashed lines indicate the leading logarithmic 
 divergence $\propto m_f \log a$ in both fits.
 The remaining $a$-dependence is consistent with lattice artifacts.
 We always use $m_s=m_s^{\rm phys}$. The lattice spacing is normalized to
 $a_0=1.46\textmd{ GeV}^{-1}$.}
\end{figure}

In the left panel of
Fig.~\ref{fig:mphys_adep} we show the $a$-dependence of the $T=0$
light quark tensor coefficient $\tau_{ub}\cdot Z_T$
at the physical point and the result of the above interpolation,
including the systematic error estimated using the different fits.
For demonstration purposes, we also indicate the leading logarithmic behavior, 
that we obtain by subtracting the lattice artifact terms from the central fit.
Comparing to the similar plot for $\chi_b$ (Fig.~\ref{fig:chi0}), we see
that deviations from the continuum behavior are sizable (and are
predominantly due to the fact that we are 
dealing with a dimensionful quantity in this case). For this reason, here
we cannot reliably determine the value of $\muqed$. Nevertheless, we note
that fixing the renormalization scale to its value from Eq.~\eqref{eq:mures}
also gives acceptable fits.
This is in agreement with the expectation of Sec.~\ref{sec:decomp1}, as well 
as with the results in the free case, see App.~\ref{sec:tensorbfree}.
The logarithmic divergence $\propto m_f\log a$ becomes more pronounced for heavy 
quarks. This is visible in the right panel of Fig.~\ref{fig:mphys_adep}, where 
we plot the strange quark tensor coefficient $\tau_{sb}\cdot Z_T$
at $m_{ud}=m_{ud}^{\rm phys}$ 
against the lattice spacing and again indicate the leading logarithmic term. 

As we have discussed above, at non-vanishing values of the
  quark mass $m_f$, the tensor coefficient diverges logarithmically.
  In Ref.~\cite{Bali:2012jv} we suggested to cancel this
  by taking the logarithmic derivative with respect to the quark mass, see also Ref.~\cite{Ioffe:1983ju}:
  \be
  f_{\gamma f}^{\perp}=\left(1-m_f\frac{\partial}{\partial m_f}\right)\tau_f\cdot Z_T\,.
  \ee
  This renormalization prescription will give identical
  results for any regulator, up to the multiplicative factor
  $Z_T$.\footnote{
  This construction will not only cancel the logarithmic divergence but also 
  any finite term $\propto m_f$. Should this be unwanted then one will
  have to accept a scheme-dependence and convert between different
  schemes in a similar way as is done for the massive chiral
  condensate, e.g., in Ref.~\cite{McNeile:2012xh}.}
Using this prescription, it turns out that
  $f_{\gamma u}^{\perp}=f_{\gamma d}^{\perp}=f_{\gamma}^{\perp}$ holds
  within statistical errors. For the strange quark we obtain:
  \be
  f_{\gamma s}^{\perp}  = -68(3)(4)\textmd{ MeV}\,.
  \label{eq:fstrange}
  \ee
  The first error includes the described variation of the fit
  while the second error reflects the uncertainty of the derivative
  with respect to $m_s$ that we indirectly determine from the
  dependence of the tensor coefficient on the light quark mass,
  following the procedure explained in Ref.~\cite{Bali:2012jv}.

  In the literature often the magnetic susceptibility of the quark
  condensate,
  \be
  X_u=\frac{\tau_{u}}{\left\langle \bar{\psi}_u\psi_u\right\rangle}\cdot \frac{Z_T}{Z_S}\,,
  \ee
  is given, rather than $f^{\perp}_{\gamma}=\tau_{u}\cdot Z_T$. Since the
  latter quantity has a smaller anomalous dimension and its value does not
  depend on a separate computation of the chiral condensate, this is
  the preferred choice for practical applications. However, for
  convenience of comparison, we shall convert it into the other convention.
  The numerical value of the quark condensate in the SU(2) chiral limit in the
  $\overline{\mathrm{MS}}$ scheme at the scale
  $\muqcd=2\textmd{ GeV}$
  reads $\langle \bar{\psi}_u\psi_u\rangle=[272(5)\textmd{ MeV}]^3$~\cite{Aoki:2019cca}. To enable a comparison with other results, below we also list
  $X_u$ at the scale $\muqcd=2\textmd{ GeV}$. Since most literature values
  refer to a low, sometimes unspecified scale, in addition we run $X_u$ as well
  as $f_{\gamma}^{\perp}$ to the scale $\muqcd=1\textmd{ GeV}$, which is
  used in most sum rule calculations, see, e.g.,
  Refs.~\cite{Balitsky:1985aq,Balitsky:1989ry,Ball:2002ps}.
  This is done, using the
  five-loop $\beta$- and quark mass anomalous dimension
  $\gamma$-functions~\cite{Baikov:2016tgj,Baikov:2014qja} and the three-loop
  $\gamma$-function of the tensor
  current~\cite{Broadhurst:1994se,Gracey:2003yr}. The results read
  \begin{align}
  X_u(2\textmd{ GeV})&=-\left[665(13)\textmd{ MeV}\right]^{-2}\,,\quad
  X_u(1\textmd{ GeV}) =-\left[542(11)\textmd{ MeV}\right]^{-2}\,,\\
  f_{\gamma}^{\perp}(1\textmd{ GeV})&=-51.1(1.6)\textmd{ MeV}\,,
  \end{align}
  where we have added all errors in quadrature, including the uncertainty
  of $f^{\perp}_{\gamma}(2\textmd{ GeV})$, the difference between running with the
  two- and three-loop $\gamma$-functions of the tensor current, the uncertainty
  of $\langle \bar{\psi}_u\psi_u\rangle$ and the uncertainty of the
  strong coupling parameter~\cite{Bruno:2017gxd}. All the above results
  are in the $\overline{\mathrm{MS}}$-scheme.

  We summarize earlier results from the literature for comparison.
  The first sum rule determination of $X_u$~\cite{Ioffe:1983ju} suggested
  a value $X_u(0.5\textmd{ GeV})=-[350(50)\textmd{ MeV}]^{-2}$ while
  vector meson dominance yields~\cite{Balitsky:1983xk} $X_u\approx
  2/m_{\rho}\approx -(540\textmd{ MeV})^{-2}$. This was improved upon in
  subsequent sum rule determinations, see, e.g., Ref.~\cite{Ball:2002ps}
  and references therein. The most extensive sum rule
  study~\cite{Ball:2002ps}
  found $X_u(1\textmd{ GeV})\approx -(560\textmd{ MeV})^{-2}$, which
  agrees reasonably well with our determination.
  A comparatively smaller absolute value
  $f_{\gamma}^{\perp}\approx -38\textmd{ MeV}$ was obtained at a low scale
  $\mu\sim 600\,\textmd{ MeV}$
  in the quark-soliton model~\cite{Petrov:1998kg} while the Vainshtein
  relation~\cite{Vainshtein:2002nv} suggests
  an even smaller modulus of the magnetic susceptibility of the quark
  condensate $X_u=-N_c/(4\pi^2F_{\pi}^2)\approx
  -(335\textmd{ MeV})^{-2}$. This
  parameter was also considered in holographic studies, with the
  result $X_u\approx -(295\textmd{ MeV})^{-2}$~\cite{Gorsky:2009ma},
  while NJL- and quark-meson-model predictions give
  $X_u\approx -(480\textmd{ MeV})^{-2}$~\cite{Frasca:2011zn} and
  $X_u\approx -(440\textmd{ MeV})^{-2}$~\cite{Frasca:2011zn},
  respectively. Finally,
  quenched lattice simulations, without renormalization, gave the values
  $X_u\approx -[804(3)\textmd{ MeV}]^{-2}$ in SU(2) gauge
  theory~\cite{Buividovich:2009ih}
  and
  $X_u\approx -[486(21)\textmd{ MeV}]^{-2}$ in SU(3)~\cite{Braguta:2010ej}.
  Our previous full QCD study~\cite{Bali:2012jv} resulted
  in $X_u(2\textmd{ GeV})=
  -[693(13)\textmd{ MeV}]^{-2}$, however, in that case the renormalization
    was only carried out perturbatively.

  Our result~\eqref{eq:fstrange} for the strange quark coefficient
  translates into
  \begin{align}
  X_s(2\textmd{ GeV})&=-\left[565(50)\textmd{ MeV}\right]^{-2}\,,\quad
  X_s(1\textmd{ GeV})=-\left[460(41)\textmd{ MeV}\right]^{-2}\,,\\
  f_{\gamma s}^{\perp}(1\textmd{ GeV})&=-76.5(5.7)\textmd{ MeV}\,,
  \label{eq:fstrange2}
  \end{align}
  where we used the ratio
  $\langle \bar{\psi}_s\psi_s\rangle/\langle \bar{\psi}_u\psi_u\rangle=
  1.08(17)$~\cite{McNeile:2012xh} for the conversion between
  $f^{\perp}_{\gamma s}$ and $X_s$. The difference between $X_s$
  and $X_u(1\textmd{ GeV})\approx -\left(475\textmd{ MeV}\right)^{-2}$ has been reported
  to be negligible in the sum rule calculations~\cite{Balitsky:1989ry}.

\subsection[The spin contribution at $T>0$]{The spin contribution at \boldmath $T>0$}
\label{sec:res_3}

\begin{figure}[th]
 \centering
\mbox{
 \includegraphics[width=8.5cm]{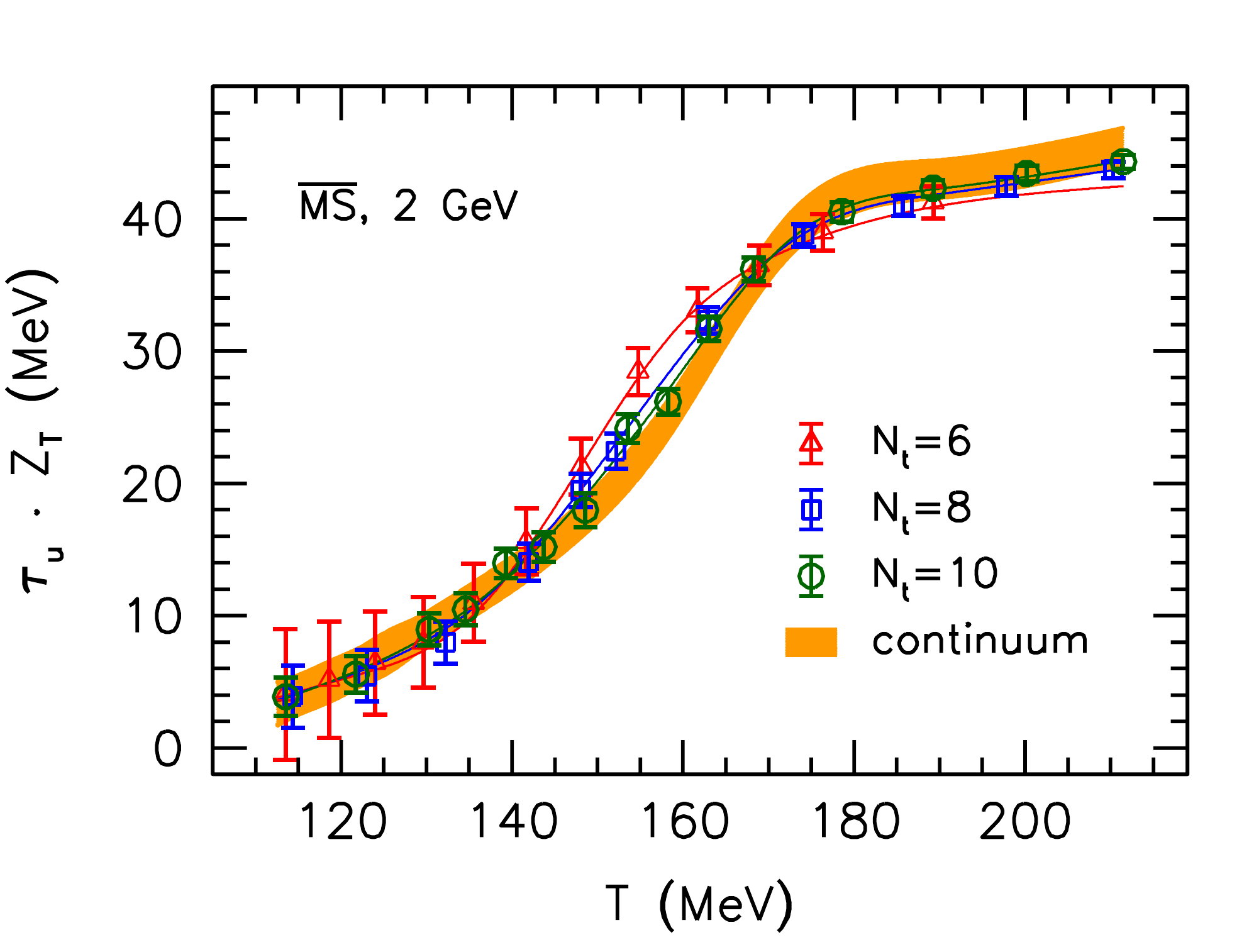}
 \includegraphics[width=8.5cm]{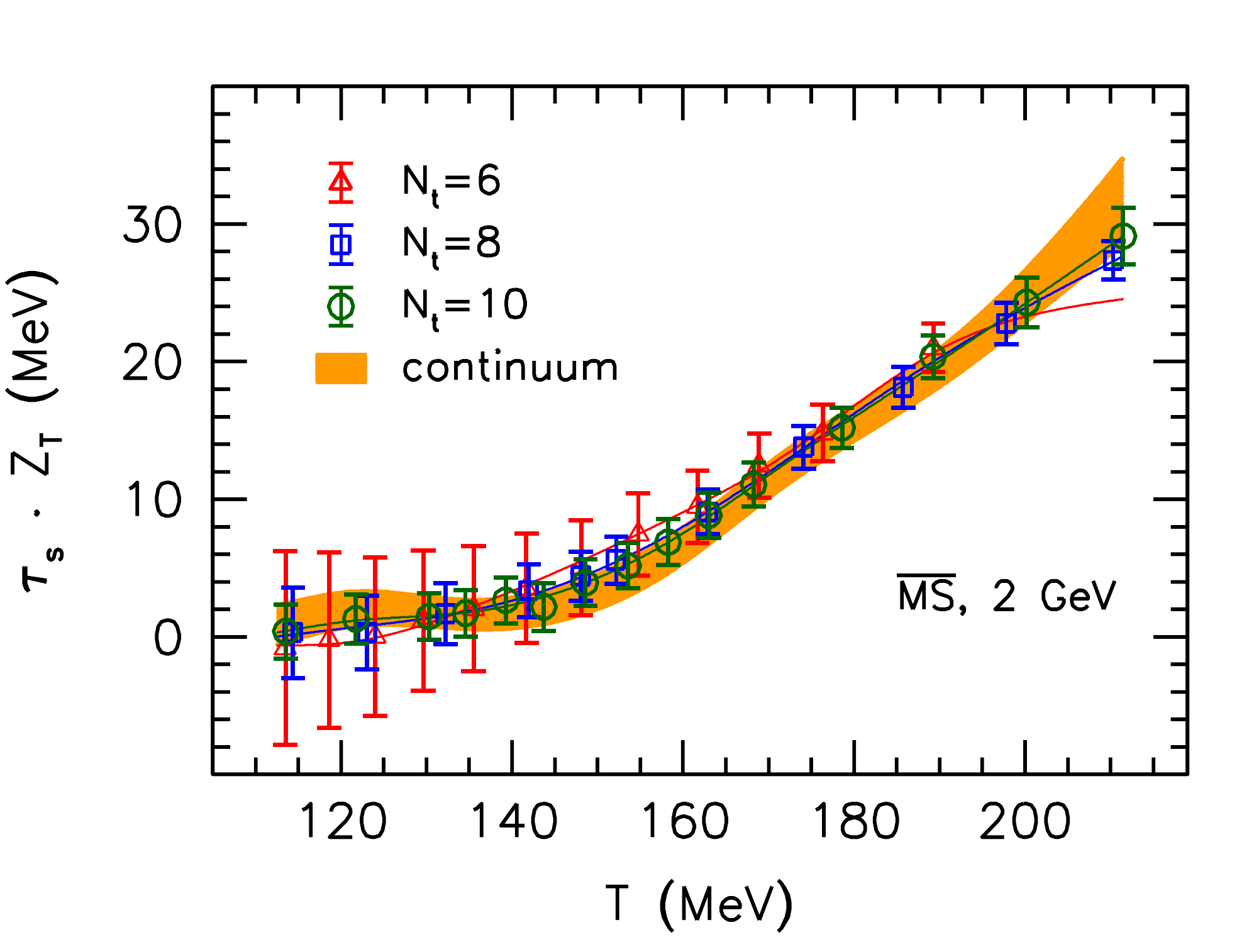}
 }
 \caption{\label{fig:finiteT}Tensor coefficients after multiplicative 
 as well as additive renormalization for the up (left panel) and for the 
 strange quark (right panel).}
\end{figure}

Having interpolated the $T=0$ tensor coefficients, we are now in the position 
to perform the additive renormalization~\eqref{eq:taufTsub} by subtracting this contribution from the finite temperature results. We use our existing $N_t=6$, $8$ and $10$ results from Ref.~\cite{Bali:2012jv}
to approach the continuum limit. Especially in light of the slow convergence 
of $\chi$ towards $a\to0$, see the right panel of Fig.~\ref{fig:artefacts},
this extrapolation should be backed up with finer lattice ensembles
in the future.
In analogy to the analysis of $\chi$, 
again we carry out a multi-spline fit of all data sets, 
determining a systematic error by varying the positions of
the spline node points.
The so-obtained fit is shown for the up quark and the strange quark in Fig.~\ref{fig:finiteT}. The results for $\tau_d$ are consistent with $\tau_u$ within errors. The large errors of our $N_t=6$ results at low temperatures are due to
the uncertainties of the $T=0$ contributions on our coarse lattices,
see Fig.~\ref{fig:mphys_adep}.

After the additive renormalization, the tensor coefficient
vanishes by definition at $T=0$. For the light quarks
$\tau_{u}(T)$ grows substantially 
as the temperature is increased, before the slope reduces and a plateau is
approached.
The inflection point of the continuum curve is found to be at $T_c=158(5) \textmd{ MeV}$. The chiral transition temperature
determined from the inflection point of the quark condensate $T_c=155(4) \textmd{ MeV}$~\cite{Borsanyi:2010bp} is in agreement
with this value. For the strange quark pseudo-critical thermal
effects set in at somewhat higher temperatures~\cite{Borsanyi:2010bp}.
Also in our case $\tau_s$ does not appear to exhibit any
inflection point, at least for $T\lesssim 170 \textmd{ MeV}$, and
below $T\approx 200 \textmd{ MeV}$ no saturation into a plateau is visible.
For sufficiently high temperatures, where the finite quark mass 
becomes negligible, we expect the two renormalized tensor coefficients 
to coincide.

Next, the continuum extrapolated results are inserted into
Eq.~\eqref{eq:spinpart1}
to determine the spin contribution $\chi^{\rm spin}$ to the susceptibility.
To this end we need to evaluate the tensor bilinear for massless valence quarks. 
Instead of performing measurements at additional valence quark masses,
we estimate this limit 
using the difference between the results for the strange quark and for the 
light quarks. 
We assume a linear dependence on the valence quark mass in the 
range $[0,m_s]$, which implies that
\be
\lim_{m^{\rm val}_{u} \to0} \tau_{u} = \lim_{m_s^{\rm val}\to0} \tau_{s}  \approx \frac{\tau_{u} m_s - \tau_{s} m_{ud}}{m_s-m_{ud}}
= \tau_{u} \frac{R}{R-1} - \tau_{s} \frac{1}{R-1} \,, \quad\quad R \equiv\frac{m_s}{m_{ud}}\,.
\label{eq:approximation}
\ee
In this case the contributions of all flavors to $\chi^{\rm spin}$ are proportional 
to $\tau_{s}-\tau_{u}$ and the renormalized spin susceptibility~\eqref{eq:spinpart1} simplifies to
\be
\chi^{\rm spin} \approx \frac{1}{2 m_{ud}} \frac{\tau_{s } - \tau_{u }}{R-1}\cdot Z_TZ_S\cdot \sum_f (q_f/e)^2\,.
\label{eq:approximation2}
\ee
Thus, in this approximation the individual flavors simply contribute in
proportion to their squared electric charges. The scalar renormalization constants
entering this expression 
are displayed in Fig.~\ref{fig:renconsts} of App.~\ref{sec:renormconstants}.

\begin{figure}[t]
 \centering
 \mbox{
 \includegraphics[width=8.5cm]{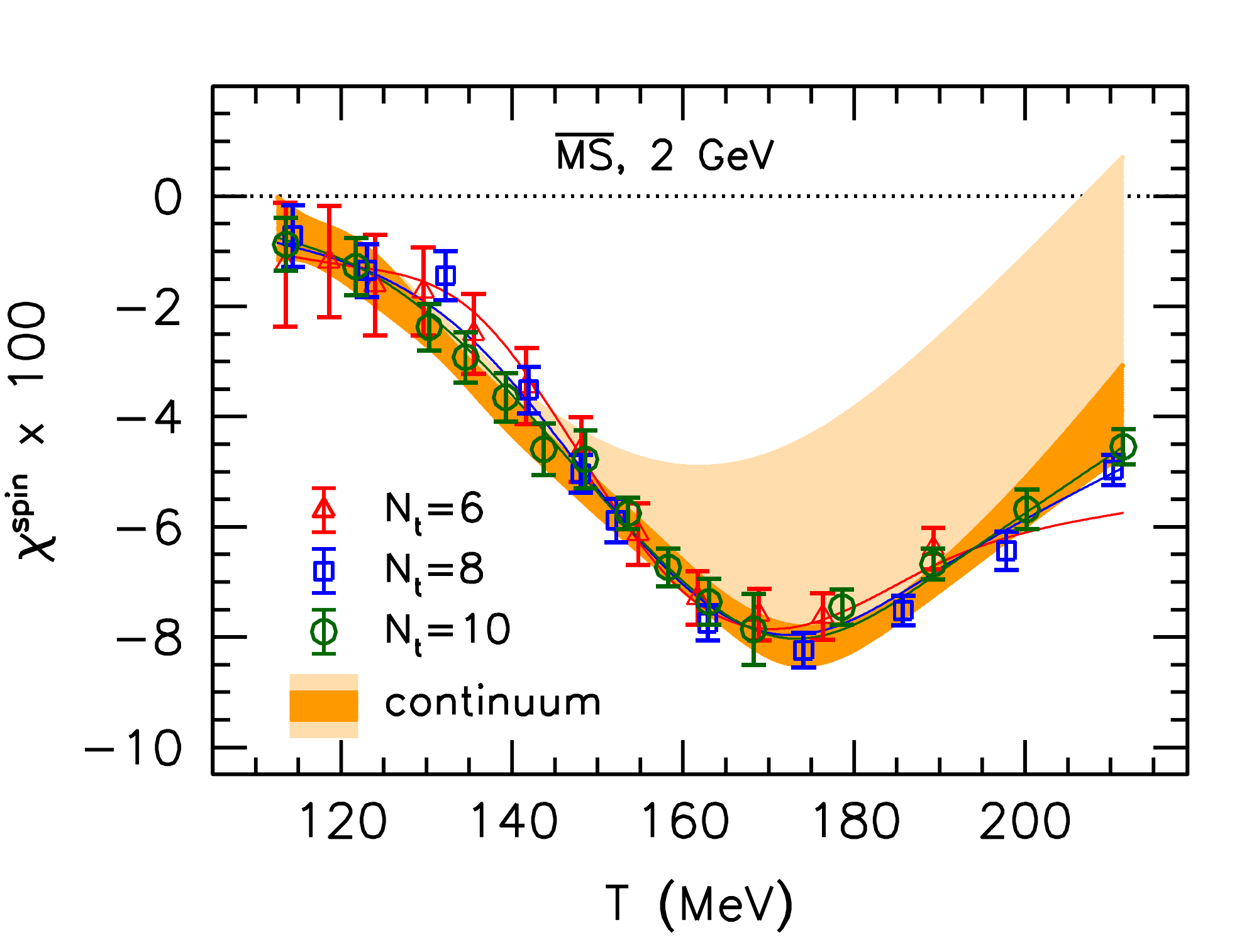}
 \includegraphics[width=8.5cm]{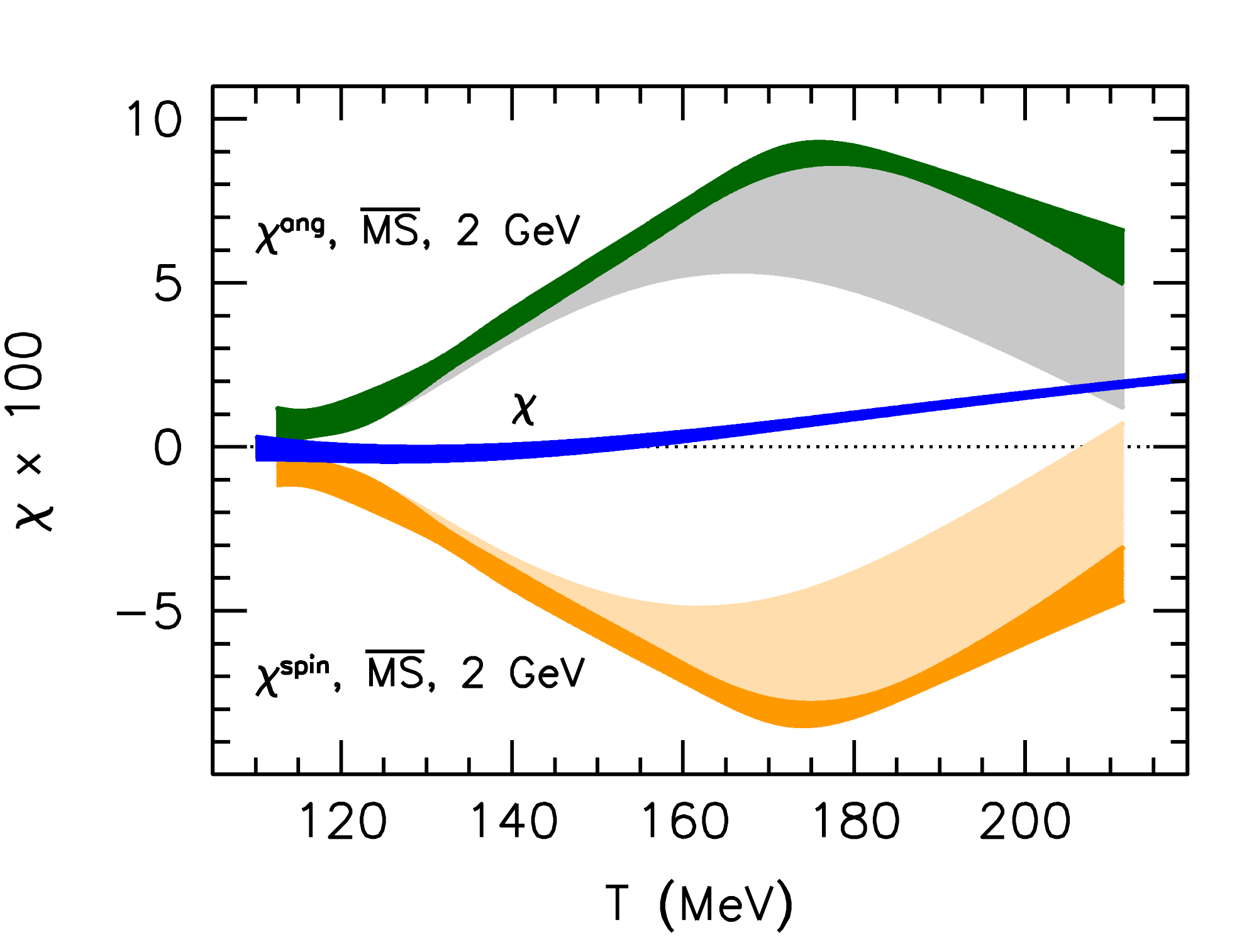}
 }
 \caption{\label{fig:chispin}Left panel: spin contribution to the susceptibility using three lattice spacings (colored symbols)
 and an extrapolation to the continuum limit (orange band). 
 A systematic uncertainty,
 related to the estimation
 of the tensor coefficient for massless valence quarks, is 
 indicated by the light yellow band. Right panel: 
 the total magnetic susceptibility from Fig.~\protect\ref{fig:chiT} (blue), together with 
 the decomposition into spin (orange-yellow)
 and orbital angular momentum (green-gray) contributions.
}
\end{figure}

The so-obtained estimate of $\chi^{\rm spin}$ is shown in 
the left panel of Fig.~\ref{fig:chispin} for three lattice spacings, together 
with a continuum extrapolation performed in the same way as for $\tau_f$.
We observe $\chi^{\rm spin}<0$ for all temperatures, 
with a
minimum somewhat above the pseudo-critical temperature and an upward 
trend for high temperatures. 
The approximation~\eqref{eq:approximation} tends to overestimate
the valence chiral limit of the tensor coefficient due to 
the presence of logarithmic deviations from a linear behavior in $m_f^{\rm val}$.\footnote{This is also visible in Fig.~\ref{fig:tauuT0}, although
  the dependencies on the valence and sea quark masses are not
  disentangled in that figure.}
Consequently, Eq.~\eqref{eq:approximation2} underestimates $\chi^{\rm spin}$.
This is also the case at high temperatures, as can be checked using the analytic formula valid for the free case, see App.~\ref{app:freecase}. To take this effect into account we include 
a systematic error based on the free case formula~\eqref{eq:highTexpansion_free}. 
In particular, we consider the difference between the approximation and the true value
in the free case
and scale it with 
the typical magnitude of the light quark tensor coefficient at lower
temperatures (see Fig.~\ref{fig:finiteT}).
The so-obtained uncertainty is also included in the 
left panel of Fig.~\ref{fig:chispin}.

We remark that $\chi^{\rm spin}<0$ for the temperature range covered
in our simulations. This can be understood by
noting that Eq.~\eqref{eq:approximation2} is the discretization of 
the mass-derivative of $\tau_f$. Increasing the mass
pushes the inflection point of $\tau_f$ to higher temperatures
(visible in Fig.~\ref{fig:finiteT}), thus making the 
derivative negative around the transition temperature.
Nevertheless, $\chi^{\rm spin}$ will necessarily turn positive
for even higher temperatures. Indeed, for sufficiently high temperatures
the difference $\tau_f=\tau_{fb}(T)-\tau_{fb}(T=0)$ will 
be dominated by the $T=0$ term, so that Eq.~\eqref{eq:approximation2}
becomes proportional to $\tau_{ub}(T=0)-\tau_{sb}(T=0)$, which is positive
for any  lattice spacing (see Fig.~\ref{fig:tauuT0}). Perturbation
theory also predicts $\chi^{\rm spin}>0$ for high temperatures,
see App.~\ref{sec:spincontribapp}.

\subsection{Pauli and Landau decomposition of the magnetic susceptibility}
\label{sec:res_4}
Finally, we compare the spin contribution to the total susceptibility
in order to learn 
about the orbital angular momentum-related contribution $\chi^{\rm ang}=\chi-\chi^{\rm spin}$. All 
three susceptibilities are included in the right panel of 
Fig.~\ref{fig:chispin}.
While the errors of the two contributions are much larger than 
that of the total susceptibility, several qualitative 
comments can be made based on this plot. 
First of all, in the complete temperature range under study, 
$\chi^{\rm spin}$ and $\chi^{\rm ang}$ have opposite signs 
and $\chi$ emerges as a result of a large cancellation between 
the two terms. As we argued above, the spin part will necessarily 
turn positive for higher temperatures, eventually 
approaching $3/2$ times the full susceptibility. 
Consequently, $\chi^{\rm ang}$ will turn negative and 
approach $-1/2\cdot \chi$. It is intriguing to observe 
that in the strongly interacting regime 
the two contributions have opposite signs than in the usual free fermion picture
according to Pauli and Landau: it is the Landau term that drives
the paramagnetic response of the QCD vacuum up to temperatures
$T\gtrsim 200\textmd{ MeV}$, while the Pauli term reduces the susceptibility
in this region. 
This unusual behavior
becomes possible due to the strong interaction, which confines 
quarks into composite hadrons and thereby fixes the relative orientation
of their spins, i.e.\ their magnetic moments. In particular, in charged
pions one of the constituent quarks is bound to anti-align its magnetic
moment with the background field in order to maintain zero total spin.
Similar effects arise for certain baryons as well. Beyond this qualitative
argument, it is difficult to anticipate the outcome of this competition
between the strong and the electromagnetic forces. Our
quantitative results reveal a peculiar interplay between 
confinement and spin physics.

To further our understanding, in principle $\chi$ can also be
decomposed into $\chi_f$ for the quark
flavors $f$ and a gluonic contribution $\chi_g$. Subtracting this
$\chi_g$ from $\chi^{\rm ang}$ will isolate the total quark orbital angular
momentum contribution $\sum_f (\chi_f-\chi_f^{\rm spin})$, in analogy to
spin decompositions~\cite{Ji:1996ek} in deep inelastic scattering that
are based on the Belinfante-Rosenfeld energy-momentum tensor, in this
case of the transverse spin.
The unrenormalized qualitative results of Ref.~\cite{Bali:2013esa}
indicate that $\chi_g\sim \chi/3$ at small temperatures. It may be
interesting to address this quantitatively in the future.

\section{Summary}

In this paper we determined the magnetic susceptibility $\chi$ of the thermal QCD 
medium via a method introduced originally for $T=0$~\cite{Bali:2015msa}, which circumvents the flux quantization problem and 
allows us to express $\chi$ in terms of $B=0$ measurements. This considerably 
reduces the measurement costs as well as systematic uncertainties compared to 
previous approaches. The susceptibility is extrapolated to the continuum limit 
for a broad range of temperatures,
making contact to the Hadron Resonance Gas (HRG) model at low $T$ as well 
as to perturbation theory at high $T$. In the confined phase we find evidence 
for a diamagnetic behavior ($\chi<0$), while for
$T\gtrsim 150\textmd{ MeV}$
we observe paramagnetism ($\chi>0$). 
Our continuum extrapolations are based on four lattice spacings and are guided by 
a generalized HRG model taking into account taste splitting (see App.~\ref{sec:B}).
A careful continuum limit is found to be essential to observe diamagnetism at low $T$ 
since this is due to light pions -- we argue that this behavior was missed in 
previous investigations
because of large lattice artifacts.

The susceptibility is decomposed into spin- ($\chi^{\rm spin}$) and orbital angular momentum-related ($\chi^{\rm ang}$)
contributions based on our previous study~\cite{Bali:2012jv}. The spin term 
is shown to be given in terms of the mass-dependence of the 
$\expv{\bar\psi\sigma_\subs\psi}$ fermion bilinear in the presence of 
a small magnetic field, see Eq.~\eqref{eq:spinpart1} and App.~\ref{app:spinpart}. 
Besides its role in this decomposition, the tensor bilinear
is related to the normalization $f_\gamma^\perp$ of the photon distribution amplitude, 
relevant for a range of phenomenological applications. We update our 
previous determination~\cite{Bali:2012jv} of the corresponding tensor 
coefficient in the chiral limit at $T=0$, by performing the multiplicative 
renormalization of $\expv{\bar\psi\sigma_\subs\psi}$ non-perturbatively on the lattice.
We obtain the value $f_\gamma^\perp=-45.4(1.5)\textmd{ MeV}$
for massless quarks, in the $\overline{\rm MS}$ scheme at
a QCD renormalization scale of $2\textmd{ GeV}$. The values of the tensor
coefficient at the physical light and strange quark masses
and at different renormalization
scales are given in Eqs.~\eqref{eq:fstrange}--\eqref{eq:fstrange2}.

At finite temperatures we performed the continuum extrapolation 
of $\chi^{\rm spin}$ and also 
determined the orbital angular momentum-related susceptibility $\chi^{\rm ang}$.
In the absence of color interactions, the two contributions exhibit the 
constant ratio $\chi^{\rm spin}:\chi^{\rm ang} = 3:(-1)$ as is well known since 
the analysis of the free electron gas by Pauli~\cite{Pauli:1927:GPG} and Landau~\cite{Landau1930}. Around the transition temperature, in full QCD
this ratio is instead found to be close to $(-1):(1.03)$, resulting
in a large cancellation between the two contributions, thereby
substantially reducing the total susceptibility.
As the temperature grows the susceptibilities approach their free-case 
counterparts, which are discussed in detail in App.~\ref{app:freecase}.
Still, it is stunning to observe that in the strongly coupled QCD medium
$\chi^{\rm spin}$ and $\chi^{\rm ang}$ have signs that are opposite to the naive expectations.

Considering our results at high temperature, it is interesting to make a comparison 
to a classical ideal system. In such a setting the Bohr-van Leeuwen theorem~\cite{Bohr,BVL} 
(for a recent review, see Ref.~\cite{BVL2}) holds: the total magnetization vanishes, since the magnetic field 
does not transfer any work to the electric currents in the system. 
Apparently, the QCD medium 
does not become classical in this sense for $T\to\infty$, even if the $\mathcal{O}(B^2)$ terms of the free energy density 
that we have discussed in this paper are small 
compared to the dominant $\mathcal{O}(T^4)$ contributions in that limit.
The non-classicality has two different origins. First, quark spins are of quantum 
nature and can induce a magnetization by aligning with the magnetic field. 
Second, both $\chi^{\rm spin}$ and $\chi^{\rm ang}$ diverge as $\log T$ for 
high temperatures. This behavior stems from the renormalization properties 
of the {\it bare} susceptibilities: quantum effects give rise to a logarithmic 
divergence $\propto\log 1/a$ in the cut-off. In turn, the 
same behavior shows up in the {\it renormalized} susceptibilities 
if they are probed by another large dimensionful scale, the temperature. 
Note that a similar connection exists
between the logarithmic divergence and the behavior of the 
renormalized free energy in the $B\to\infty$ limit~\cite{Dunne:2004nc}.

\acknowledgments
This research was funded by the DFG (Emmy Noether Program EN 1064/2-1 and SFB/TRR 55). The authors would like to thank Falk Bruckmann and V.~M.~Braun for
enlightening discussions, Nikolay Kivel and
Massimo D'Elia for insightful comments as well as 
Ren-Hong Fang for pointing out a mistake in an earlier version of 
App.~\ref{sec:highTexp}.

\appendix

\section{The HRG model and lattice discretization errors}
\label{sec:B}

At low temperatures the staggered action suffers from enhanced lattice artifacts due to taste splitting. 
Here we attempt to incorporate the effects of this splitting into the HRG model. 
The magnetic susceptibility was calculated in a standard HRG model in Ref.~\cite{Bali:2014kia}.
Following Ref.~\cite{Huovinen:2009yb} we replace the contribution of pions in the model by a sum over each taste, weighted by 
the corresponding degeneracies. The masses of the individual tastes and their parametrization 
in the range of our lattice spacings are taken from Ref.~\cite{Borsanyi:2010cj}. Since pions are dominant for the susceptibility, the taste splitting for other mesonic and baryonic states is 
ignored for simplicity (although the splitting for $\eta$ mesons might also lead to light mesonic states, see, e.g., Ref.~\cite{Bazavov:2009bb}). The list of hadrons taken into account can be found 
in Ref.~\cite{Endrodi:2013cs}.

\begin{figure}[t]
 \centering
 \includegraphics[width=8.5cm]{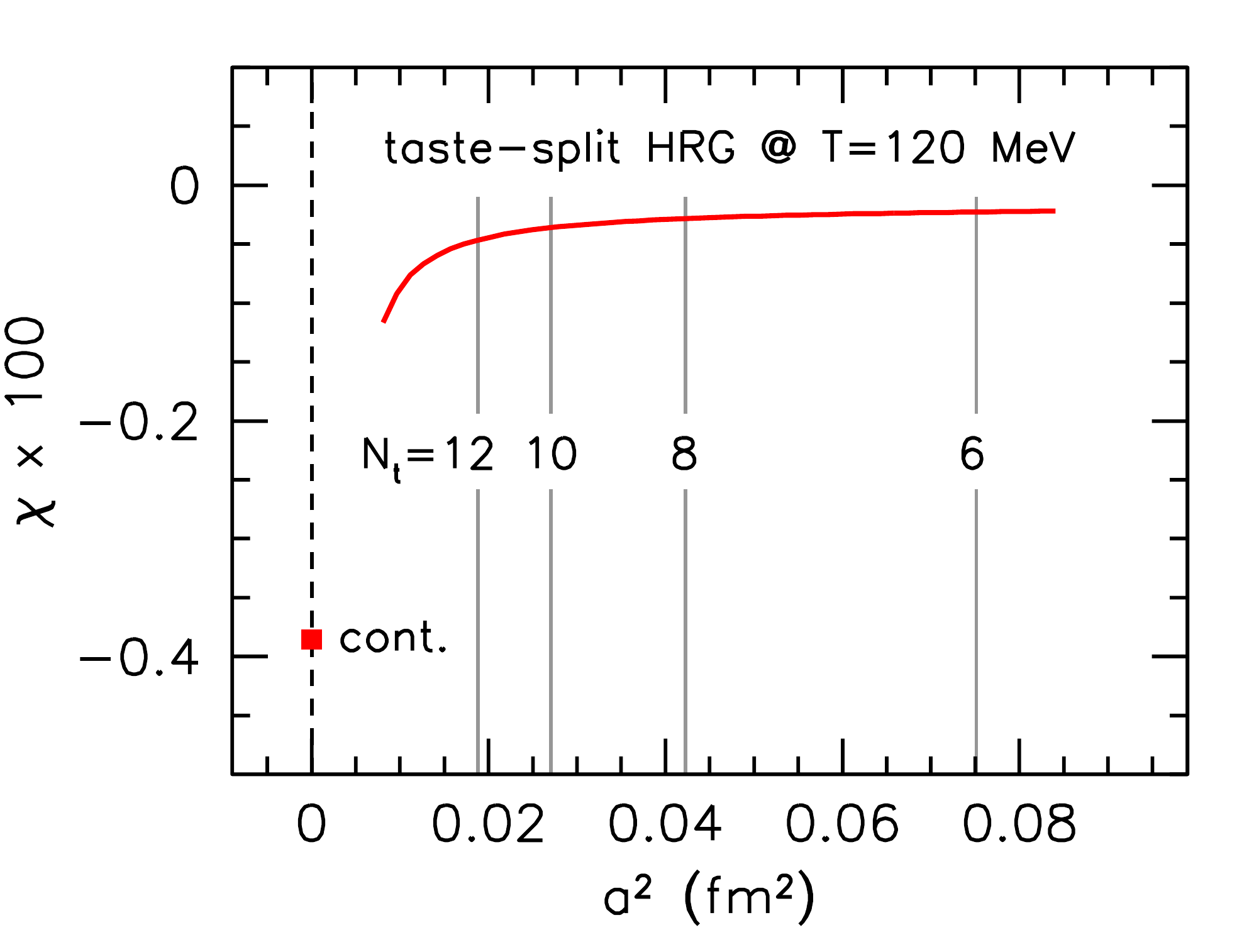}
 \caption{\label{fig:chiT_hrg} Lattice artifacts in the susceptibility in a generalized HRG model involving taste splitting.}
\end{figure}

In Fig.~\ref{fig:chiT_hrg} 
we show the renormalized magnetic susceptibility evaluated at $T=120\textmd{ MeV}$ as a function of the lattice 
spacing. The spacings for our four ensembles $N_t=6,8,10$ and $12$ at this temperature 
are highlighted in the plot. This reveals slow convergence towards the continuum limit, which
can best be understood by analyzing the mass-dependence of the pionic contribution $\chi_\pi$ to the susceptibility, which takes the form~\cite{Bali:2014kia}
\be
\chi_\pi(m_\pi) =
-\frac{1}{48\pi^2} \int_0^\infty \frac{\dd t}{t} e^{-m_\pi^2t/T^2}
\left[ \Theta_3\left(0, e^{-1/(4t)}\right)-1 \right] \,,
\ee
where $\Theta_3$ is an elliptic $\Theta$-function.
This can be derived by comparing to the analogous expression for fermions, calculated below in Eq.~\eqref{eq:C8}. The bosonic Matsubara frequencies 
give rise to the different first argument in the elliptic function. The prefactor 
in this case is the scalar QED $\beta$-function coefficient 
for one complex scalar field $\beta_1^{\rm scalar} = 1/(48\pi^2)$.
The pionic susceptibility diverges logarithmically in the chiral limit
(this can be shown similarly to the calculation below in App.~\ref{sec:highTexp}),
\be
\chi_\pi(m_\pi) \xrightarrow{m_\pi\to0} -\beta_1^{\rm scalar} \log (T/m_\pi)^2 \,,
\ee
explaining its pronounced dependence on $m_\pi$. In turn, nonzero lattice spacings enhance the masses of 
most pion tastes, thus, reducing the magnitude of $\chi_\pi$.

Based on the HRG predictions for $\chi(a,T)$ we consider the difference between 
a simple $\mathcal{O}(a^2)$ fit taking into account only $N_t\le12$ lattices and 
the true continuum limit. This difference is included as a lower systematic error 
of our lattice determination of $\chi(T)$ at low temperatures, see Fig.~\ref{fig:chiT}.

\section{Separation into quark spin and other angular momentum contributions}
\label{app:spinpart}

Here we derive the relation between the spin contribution to the susceptibility
and the tensor bilinear, as shown in Eqs.~\eqref{eq:taufbdefinition}--\eqref{eq:spinpart1} of the main text.
It is instructive to begin with the first derivative of the free energy 
density,
\be
-\frac{\partial f}{\partial B}=\frac{T}{V}\sum_f\expv{ \tr \frac{1}{\dsf+m_f}\frac{\partial \dsf}{\partial B}}
= \frac{T}{2V}\sum_f\expv{ \tr \frac{1}{(\dsf+m_f)\dsf}\frac{\partial \dsf^2}{\partial B}}\,,
\label{eq:C1}
\ee
where we used the cyclicity of the trace (even though $\dsf$ and $\partial\dsf/\partial B$ do not commute, we can symmetrize the expression in the
two operators under the trace).
Now we use the relation
\be
\frac{1}{(\dsf+m_f) \dsf} = -\frac{1}{m_f}\left[\frac{1}{\dsf+m_f} - \frac{1}{\dsf}\right] \,,
\label{eq:rel1}
\ee
and the identities
\be
\frac{\partial \dsf^2}{\partial (q_fB)}
= -\sigma_\subs -L_\subs, \quad\quad
\sigma_\subs = \frac{1}{2i} [\gamma_1,\gamma_2], \quad\quad
L_\subs=-\frac{\partial D_f^2}{\partial (q_fB)}\,,
\label{eq:rel2}
\ee
where $\sigma_\subs$ is the relevant component of the relativistic spin operator 
defined in Eq.~\eqref{eq:sxydef} and $L_\subs$ is a
generalized angular momentum operator, which depends on the electromagnetic 
as well as the $\mathrm{SU}(3)$ gauge.

Using Eqs.~\eqref{eq:rel1} and~\eqref{eq:rel2}, 
we can rewrite Eq.~\eqref{eq:C1} as
\be
-\frac{\partial f}{\partial B}
= \frac{T}{2V}\sum_f\frac{q_f}{m_f} \expv{ \tr \frac{\sigma_\subs+L_\subs}{\dsf+m_f} - 
\tr \frac{\sigma_\subs+L_\subs}{\dsf}}
= \sum_f\frac{q_f}{2m_f} \bigg[1-\lim_{m_f^{\rm val}\to0}\bigg]\expv{ \bar\psi_f \sigma_\subs \psi_f + \bar\psi_f L_\subs\psi_f}
\,.
\label{eq:C4}
\ee
Thus, in the language of Eq.~\eqref{eq:Zseaval}, 
we need the difference of two terms: one with valence quark mass
$m_f^{\rm val}=m_f$ and one with $m_f^{\rm val}\to0$. The sea quark 
mass is kept fixed in both cases: $m_f^{\rm sea}=m_f$.
We remark that the vanishing valence quark mass needs
to be defined as a limit in finite volumes (see below). 
Also note that the fermion bilinears are defined to include the volume factor $T/V$.

Differentiating Eq.~\eqref{eq:C4} once more with respect to $B$ at $B=0$
and dividing by $e^2$, we recover the bare magnetic susceptibility~\eqref{eq:defchi} on the left hand side,
\be
\chi_b = \sum_f \frac{(q_f/e)^2}{2m_f} \bigg[1-\lim_{m_f^{\rm val}\to0}\bigg] \lim_{B\to0} \frac{\expv{ \bar\psi_f \sigma_\subs \psi_f + \bar\psi_f L_\subs\psi_f}}{q_fB}\,.
\label{eq:B5a}
\ee
The slope of the tensor bilinear $\expv{\bar\psi_f\sigma_\subs\psi_f}$ 
for small values of $B$ gives the 
tensor coefficient $\tau_{fb}$ as defined in Eq.~\eqref{eq:taufbdefinition}. 
After subtracting its value at $T=0$ and multiplying by the relevant QCD renormalization 
factors, this term gives the spin contribution to the susceptibility $\chi^{\rm spin}$, as we wrote in the main text, Eq.~\eqref{eq:spinpart1}. In turn, the magnetic field-dependence of the bilinear 
involving the generalized angular momentum operator $L_\subs$ 
is related to $\chi^{\rm ang}$. The latter term cannot be implemented 
straightforwardly due to its gauge-dependence and magnetic flux quantization.

In Ref.~\cite{Bali:2012jv} we already discussed the separation of the magnetic 
susceptibility into quark spin- and other angular momentum-related contributions. There, the $m_f^{\rm val}=0$ term of Eq.~\eqref{eq:C4} was argued not to contribute -- indeed, in a finite volume the massless limit of fermion bilinears always vanishes. 
However, in the thermodynamic limit this is not the case if chiral 
symmetry is broken spontaneously. 
To elucidate this point in more detail, let us rewrite the trace in Eq.~\eqref{eq:C4} using the eigenmodes of the Dirac operator,
\be
\dsf \chi_{f\lambda} = i\lambda \chi_{f\lambda}\,,
\ee
so that, exploiting chiral symmetry $\{\gamma_5,\dsf\}=0$, 
\begin{align}
\frac{T}{V}\expv{\!\tr \frac{\sigma_\subs}{\dsf+m_f^{\rm val}}\!} &= 
\frac{T\,m_f^{\rm val}}{V}\expv{\!\tr \frac{\sigma_\subs}{-\dsf^2+(m_f^{\rm val})^2}\!\!}\nonumber\\&\xrightarrow{V\to\infty}
\int_0^\infty \!\!\!\dd \lambda \,
\frac{2m_f^{\rm val}}{\lambda^2+(m_f^{\rm val})^2} \expv{\!\rho_f(\lambda;m_f^{\rm sea})\,\chi_{f\lambda}^\dagger \sigma_\subs \chi_{f\lambda}}\,,
\label{eq:C3}
\end{align}
where $\rho_f(\lambda;m_f^{\rm sea})$ is the spectral density of $\dsf$ in the infinite volume, determined in an ensemble generated with sea quark masses $m_f^{\rm sea}$.
Towards the valence chiral limit the kernel becomes proportional to the $\delta$-distribution, 
so that 
we have a Banks-Casher-type~\cite{Banks:1979yr} relation,
\be
\expv{\bar\psi_f\sigma_\subs\psi_f}
\xrightarrow{V\to\infty, \,m_f^{\rm val}\to0} \pi
\int_0^\infty \!\!\dd \lambda \,
\delta(\lambda) \expv{\rho_f(\lambda;m_f^{\rm sea})\,\chi_{f\lambda}^\dagger \sigma_\subs \chi_{f\lambda}}
=\pi \expv{\rho_f(0;m_f^{\rm sea})\,  \chi^\dagger_{f0} \sigma_\subs \chi_{f0}}\,.
\ee
On the one hand, this limit is zero if chiral symmetry is intact and the spectral 
density vanishes at the origin. On the other hand,
a nonzero chiral condensate $\expv{\rho_f(0;m_f^{\rm sea})}$, together with the 
polarization $\sigma_\subs\chi_{f0}=\chi_{f0}$ of the low modes
will turn the chiral limit of the tensor bilinear nonzero.
Our lattice results reveal a nonzero value for $\expv{\bar\psi_f\sigma_\subs\psi_f}$ in the full chiral limit at low temperatures, see Fig.~\ref{fig:tauuT0}. Clearly, the fermion bilinear remains nonzero also if 
only $m_f^{\rm val}$ is sent to zero.
This is in accordance with the recent findings of Ref.~\cite{Bruckmann:2017pft} about the Dirac spectrum 
at $B>0$, where the low modes were indeed found to exhibit 
almost perfect spin-polarization.

\section{Susceptibilities in the free case}
\label{app:freecase}

Here we consider the free case (i.e.\ we set the color charges of quarks
to zero) to exemplify the most important relations of the main text. 
These include the proportionality between the tensor bilinear and 
the spin contribution to the susceptibility, the 
ultraviolet divergences of the susceptibilities 
at zero temperature as well as the high-temperature behavior of the 
renormalized susceptibilities.
These calculations include our previous results~\cite{Bali:2012jv,Bali:2014kia}, which we also show here for completeness.

Below we will extensively use Schwinger's proper time formulation~\cite{Schwinger:1951nm}. This is based on the Mellin transform
\be
E^{-z} = \frac{1}{\Gamma(z/2)} \int_0^\infty \dd t \,t^{z/2-1} \,e^{-E^2 t}\,,
\label{eq:MELLIN}
\ee
and its inverse
\be
e^{-lE/T} = \frac{1}{2\pi i} \int_{c-i\infty}^{c+i\infty} \dd z \,\Gamma(z) \,l^{-z} E^{-z} \,T^z\,,
\label{eq:MELLININV}
\ee
which are valid for $\textmd{Re} \,z>0$, $c>0$ and $E>0$. 
Moreover, taking the derivative of Eq.~\eqref{eq:MELLIN} with respect to $z$ 
at $z=0$ gives the standard $\zeta$-function regularization 
result~\cite{Elizalde:1994gf},
\be
\log E^2 = -2\left.\frac{\partial \,(E^2)^{-z/2}}{\partial
z}\right|_{z=0} 
= -2\left.\frac{\partial}{\partial z}\right|_{z=0}
\frac{1}{\Gamma(z/2)} \int_0^\infty \dd t \,t^{z/2-1} \,e^{-E^2 t}\,.
\label{eq:MELLIN2}
\ee

\subsection{Magnetic susceptibility}

We consider one quark flavor $\psi$ with electric charge $q$ 
and mass $m$ in a volume $V=L^3$ at temperature $T$, exposed to a background magnetic field $B$. For convenience we assume that the magnetic field is 
oriented in the $x_3$ direction and $qB>0$.
The free energy density in this setting reads (see, e.g., Ref.~\cite{Fraga:2012rr}):
\be
f(B,T)=- N_c\,\frac{qB}{2\pi} \sum_{k=0}^\infty \sum_{s=\pm1/2} T\sum_{n=-\infty}^\infty \int\frac{\dd p}{2\pi}
\log\frac{\omega_n^2 + E_{p,s,k}^2}{T^2}\,,
\label{eq:freeenergyT}
\ee
where $\omega_n=(2n+1)\pi T$ is the $n$-th fermionic Matsubara frequency. Moreover, $p$, $s$ and $k$ are the momentum, spin and angular momentum in the direction of the magnetic field,
$N_c=3$ is the number of colors 
and the energies are given by the Landau levels,
\be
E_{p,s,k} = \sqrt{p^2+m^2+(2k+1-2s)qB}\,.
\label{eq:energies}
\ee

Rewriting the logarithm using Eq.~\eqref{eq:MELLIN2}, the integral over $p$ becomes Gaussian and can be solved. Furthermore, 
the sums over $n$, $k$ and $s$ are
\begin{align}
T\!\sum_{n=-\infty}^{\infty} \!\!e^{-\omega_n^2 t} = \frac{1}{2\sqrt{\pi t}} \,\Theta_3\!
\left(\frac{\pi}{2},e^{-1/(4tT^2)}\right)&,\;\,
\sum_{k=0}^{\infty} e^{-(2k+1)qB\,t} = \frac{1}{2\sinh(qBt)},\nonumber\\
\sum_{s=\pm1/2} \!\!e^{-2sqB\,t} &= 2\cosh(qBt)\,,
\label{eq:freesums}
\end{align}
where $\Theta_3$ is an elliptic function. 
Inserting these in Eq.~\eqref{eq:freeenergyT} and performing the derivative with 
respect to $z$, we obtain
\be
f(B,T)= N_c \,\frac{qB}{8\pi^2} \int_0^\infty \frac{\dd t}{t^2} \, e^{-m^2t}\, \coth(qBt)\;\Theta_3\!
\left(\frac{\pi}{2},e^{-1/(4tT^2)}\right)\,.
\label{eq:C7}
\ee
Taking the second derivative with respect to $eB$ to obtain the 
bare magnetic susceptibility~\eqref{eq:defchi} results in
\be
\chi_b(T) = -\frac{N_c}{12\pi^2} (q/e)^2 \int_{0}^\infty \frac{\dd t}{t} e^{-m^2t}\,\; \Theta_3\!
\left(\frac{\pi}{2},e^{-1/(4tT^2)}\right)\,.
\label{eq:C8}
\ee

\subsection{Ultraviolet divergences and QED renormalization}
\label{sec:UVdivs}

To determine the ultraviolet structure of the magnetic susceptibility, 
we consider Eq.~\eqref{eq:C8} at zero temperature. 
For $T=0$ the elliptic function $\Theta_3$ approaches unity. 
The resulting expression needs to be regularized, for example by setting an ultraviolet cut-off $1/\Lambda^2$ as the lower limit of the proper time integral. Performing the integral and expanding for large $\Lambda$ we obtain,
\be
\chi_b(T=0)= \frac{N_c}{12\pi^2} (q/e)^2 \left[ \log\frac{\Lambda^2}{m^2} -\gamma_E\right] + \mathcal{O}(\Lambda^{-2})\,,
\label{eq:C16}
\ee
where $\gamma_E$ is the Euler-Mascheroni constant. 
Thus, the coefficient of the logarithmic divergence  indeed equals the lowest-order QED $\beta$-function coefficient $\beta_1$ (for one quark flavor with electric charge $q$), demonstrating the validity of Eq.~\eqref{eq:b1}. In fact, this relation 
continues to hold in full QCD as well, owing to the fact that towards the continuum 
limit QCD corrections to $\beta_1$ at the scale $1/a$ approach zero due to asymptotic freedom (see Eq.~\eqref{eq:formula}).
We note moreover that in the proper time formulation the renormalization scale 
is set by the mass -- in fact $\muqed=m \,e^{\gamma_E/2}$ for our choice of
the regulator $\Lambda$ -- explaining the appearance of $m$ 
in the argument of the logarithm in Eq.~\eqref{eq:C16}.

The additive renormalization can be performed by subtracting $\chi_b(T=0)$ from Eq.~\eqref{eq:C8}:
\be
\chi = \chi_b(T)-\chi_b(T=0) = 
-\frac{N_c}{12\pi^2} (q/e)^2 \int_{0}^\infty \frac{\dd t}{t} e^{-m^2t}\,\; \left[ \Theta_3\!
\left(\frac{\pi}{2},e^{-1/(4tT^2)}\right) -1 \right]\,.
\label{eq:renormchifree}
\ee
As we mentioned after Eq.~\eqref{eq:renorm}, this corresponds to the choice of a physical, albeit scheme-dependent,
QED renormalization scale.

\subsection{Spin contribution}
\label{sec:spincontribapp}

The contribution of orbital angular momentum to the total susceptibility can 
be calculated by simply replacing the fermion with two ghost particles (spin-zero but antiperiodic in Euclidean time) in the above calculation.
This removes the $-2sqB$ from the energies~\eqref{eq:energies}
and excludes the spin sum in the free energy density~\eqref{eq:freeenergyT}.
Consequently, 
the magnetic field-dependent part in Eq.~\eqref{eq:C7} changes 
as $\coth(qBt) \mapsto 1/\sinh(qBt)$. This merely changes the second derivative 
of the free energy density
at $B=0$ by a factor $-1/2$. Thus, for the renormalized susceptibility we arrive at
\be
\chi^{\rm ang}(T) = -\frac{1}{2} \cdot \chi(T)\,,
\ee
which also implies
\be
\chi^{\rm spin}(T) = \frac{3}{2} \cdot \chi(T)\,,
\label{eq:C10}
\ee
confirming the $3:(-1)$ ratio of the two contributions to the total susceptibility.
We mention that a similar argument has been used in perturbative QCD (with chromomagnetic background fields)
to relate asymptotic freedom to spin effects~\cite{Nielsen:1980sx}. 

\subsection{Tensor bilinear}
\label{sec:tensorbfree}

For the tensor bilinear we begin with the result of the fermionic path integral,
\be
\expv{\bar\psi \sigma_\subs\psi} = 
\frac{T}{V}\,\tr\frac{\sigma_\subs}{\slashed{D}+m} = \frac{T\,m}{V}\,\tr \frac{\sigma_\subs}{-\slashed{D}^2+m^2}\,,
\ee
where we used chiral symmetry $\{\gamma_5,\slashed{D}\}=0$.
The trace is represented using the eigenbasis of $-\slashed{D}^2$, giving the 
eigenvalues $\omega_n^2+E_{p,s,k}^2$. Since $[\slashed{D}^2,\sigma_\subs]=0$, 
the spin operator can also be diagonalized in this basis and its eigenvalues 
are minus two times the spin: $\sigma_\subs\to-2s$. Taking into account the
$2N_c \cdot(qBL^2)/(2\pi)$-fold degeneracy of the eigenvalues, we obtain
\be
\expv{\bar\psi\sigma_\subs\psi} = 
N_c \frac{qB \,m}{\pi} \sum_{k=0}^\infty \sum_{s=\pm1/2} T\sum_{n=-\infty}^\infty \int \frac{\dd p}{2\pi} \frac{-2s}{\omega_n^2+p^2+m^2 + (2k+1-2s)qB}\,.
\ee
In the sum the contributions $\{k, s = 1/2\}$
and $\{k+1, s=-1/2\}$ cancel, leaving only the
unpaired lowest Landau level $\{k = 0, s = 1/2\}$. Hence we get
\be
\expv{\bar\psi\sigma_\subs\psi} = 
-N_c \frac{qB\, m}{\pi} \,T\sum_{n=-\infty}^\infty\int \frac{\dd p}{2\pi} \frac{1}{\omega_n^2+p^2+m^2}\,.
\label{eq:C15a}
\ee
Note that, unlike in full QCD, here the tensor bilinear is exactly linear in the 
magnetic field. Thus, the tensor coefficient $\tau_b$ of Eq.~\eqref{eq:taufbdefinition} is obtained by simply dividing Eq.~\eqref{eq:C15a} by $qB$.

Using Eq.~\eqref{eq:MELLIN} with $E=\sqrt{\omega_n^2+p^2+m^2}$, performing the 
Gaussian integral over $p$ and the Matsubara sum~\eqref{eq:freesums} over $\omega_n$, we arrive at
\be
\tau_b(T) = -N_c\frac{m}{4\pi^2} \int_0^\infty \frac{\dd t}{t} \, e^{-m^2t}\;\Theta_3\!
\left(\frac{\pi}{2},e^{-1/(4tT^2)}\right)\,.
\label{eq:C14}
\ee
A comparison to Eq.~\eqref{eq:C8} reveals that this quantity contains the same logarithmic 
divergence as $\chi_b$, just with a different coefficient. Using a cut-off regulator as 
in Eq.~\eqref{eq:C16}, we obtain at $T=0$,
\be
\tau_b(T=0) = 
 \frac{N_c}{4\pi^2}\, m \left[ \log\frac{\Lambda^2}{m^2} -\gamma_E\right] + \mathcal{O}(\Lambda^{-2})\,,
\label{eq:taubdivfree}
\ee
confirming Eq.~\eqref{eq:t1}. The same considerations regarding QCD corrections to the coefficient
and the renormalization scale $\muqed$ apply as in Sec.~\ref{sec:UVdivs} for $\chi_b$.

The difference $\tau = \tau_b(T)-\tau_b(T=0)$ is ultraviolet-finite. We can compare this with Eqs.~\eqref{eq:renormchifree} and~\eqref{eq:C10} to conclude that
\be
 \frac{(q/e)^2}{2m} \left[\tau(m) - \tau(m\to0) \right] = \chi^{\rm spin}\,,
\ee
confirming the relation~\eqref{eq:spinpart1} and Eq.~\eqref{eq:C10}.
Notice that $\tau$ vanishes for $m\to0$, so 
the subtraction of the massless limit is irrelevant in the free case 
(but it is relevant for the interacting system with spontaneous chiral symmetry breaking, see App.~\ref{app:spinpart}). 

\subsection{High-temperature expansion}
\label{sec:highTexp}

The temperature-dependent part of the free energy density~\eqref{eq:freeenergyT}
can be simplified using the well-known trick~\cite{kapusta2006finite} of differentiating and subsequently integrating the integrand with respect to $E_{p,s,k}$.
The result is
\be
f(B,T)-f(B,0) = -2N_c \frac{qB}{2\pi} \sum_{k=0}^\infty \sum_{s=\pm1/2}  \int\frac{\dd p}{2\pi} \,T\log\left[ 1+e^{-E_{p,s,k}/T}\right]\,.
\label{eq:FREEE}
\ee
The energy levels are given in Eq.~\eqref{eq:energies} above.
To obtain the high-temperature expansion in a closed form, we need to 
replace the logarithm by its series expansion
\be
\log(1+x)=-\sum_{l=1}^\infty\frac{(-x)^l}{l}\,.
\label{eq:LOGEXP}
\ee
This approach was used, e.g., in Ref.~\cite{Toms:1996dg} for scalars at nonzero chemical potential.

Inserting the expansion~\eqref{eq:LOGEXP} into~\eqref{eq:FREEE} and
rewriting the exponentials using Eq.~\eqref{eq:MELLIN} results in
\be
f(B,T)-f(B,0) = N_c\frac{qB\,T}{2\pi^2} \sum_{l=1}^\infty \frac{(-1)^l}{l} \sum_{k=0}^\infty \sum_{s=\pm1/2} \int\dd p \,\frac{1}{2\pi i}
\int_{c-i\infty}^{c+i\infty} \!\!\dd z \,\Gamma(z) \, l^{-z}  E_{p,s,k}^{-z}\, T^z\,.
\ee
Inserting the Mellin transform~\eqref{eq:MELLIN} for $E_{p,s,k}^{-z}$ renders the integral over $p$ Gaussian. We can reuse the 
angular momentum and spin-sums from Eq.~\eqref{eq:freesums}, giving
\be
f(B,T)-f(B,0) = N_c\frac{qB }{2\pi^{3/2}} \,\frac{1}{2\pi i} \int_{c-i\infty}^{c+i\infty} \!\!\!\dd z \,\frac{\Gamma(z)}{\Gamma(z/2)} \,T^{z+1} \sum_{l=1}^\infty \frac{(-1)^l}{l^{1+z}} 
\int_0^\infty \!\!\dd t \,t^{(z-3)/2} \,e^{-m^2 t}\, \coth(qBt)\,.
\ee

Differentiating 
the above expression twice with respect to $eB$ at $B=0$ gives
(minus) the renormalized magnetic susceptibility $\chi$. The integral over
$t$ can 
be solved via the Mellin transform~\eqref{eq:MELLIN} and gives a $\Gamma$-function,
while the sum over $l$ results in a $\zeta$-function:
\be
\sum_{l=1}^\infty \frac{(-1)^l}{l^{1+z}} = \zeta(1+z) \cdot (2^{-z}-1)\,.
\ee
Using the duplication formula~\cite{NIST:DLMF} for the ratio of $\Gamma$-functions, we arrive at
\be
\chi = -\frac{N_c}{6\pi^2 m}  (q/e)^2 \frac{1}{2\pi i} \int_{c-i\infty}^{c+i\infty} \dd z\, \Gamma\left(\frac{z+1}{2}\right) \Gamma\left(\frac{z+1}{2}\right)\zeta(1+z) \,(1-2^z) \,m^{-z}\, T^{z+1} \,.
\ee
For the validity of the Mellin transforms we needed to assume $c>0$ (as well as $m>0$). The final integral over $z$ can be solved using Cauchy's theorem, closing 
the integral towards the left and calculating the residue at 
the poles. There is a double pole at $z=-1$ and simple poles at $z=-3,-5,\ldots$. 
These $z$-values set the powers of $T$ that 
appear in the high-temperature 
expansion. Keeping the leading terms (i.e.\ $z=-1$ and $z=-3$), we finally obtain,
\be
\chi = \frac{N_c}{12\pi^2}(q/e)^2 \left[ \log\frac{T^2\pi^2}{m^2} - 2\gamma_E +
  \frac{7 \,\zeta(3)}{4 \pi^2} \,\frac{m^2}{T^2}\right]
 +
\mathcal{O}(m^4/T^4)\,,
\label{eq:highTexpansion_free}
\ee
reproducing the results of Ref.~\cite{Cangemi:1996tp}.
Notice that the coefficient of the leading logarithmic term is equal to $\beta_1$
(for one flavor with electric charge $q$), confirming Eq.~\eqref{eq:highTchi}, in agreement with Refs.~\cite{Loewe:1991mn,Elmfors:1993bm,Gies:1999xn}.
As we have seen below Eq.~\eqref{eq:C16}, in the proper time formulation
the renormalization scale is intrinsically set by the quark mass,
$\muqed=m\, e^{\gamma_E/2}$. We may express the square bracket in the leading
term as $\log (\gamma \,T^2/\muqed^2)$ with $\gamma=\pi^2\, e^{-\gamma_E}$.
Clearly, $\gamma=\mathcal{O}(1)$ depends on the definition of the regulator.
The general form is again expected to hold in full QCD~\cite{Bali:2014kia}:
in this case
QCD corrections at scales $T\gg\muqed$ are small due to asymptotic freedom.

\section{Multiplicative QCD renormalization}
\label{sec:renormconstants}
Since lattice perturbation theory is slowly convergent and high-loop results
are unavailable, we first match the local lattice QCD operators of
interest non-perturbatively to the regulator independent
RI'-MOM scheme~\cite{Martinelli:1994ty,Chetyrkin:1999pq}
and subsequently translate the result at three-loop order~\cite{Gracey:2003yr}
to the $\overline{\rm MS}$ scheme.

The quark bilinear operators are renormalized by computing
the corresponding amputated flavor non-singlet vertex functions
for different momenta on Landau gauge-fixed ensembles.
We wish to renormalize light- and strange-quark bilinears,
which can be written as linear combinations of the diagonal SU(3)
flavor-octet and -singlet currents.
In continuum schemes,
with the exception of the axial current that we do not discuss here,
the renormalization of flavor singlet and non-singlet 
operators of dimension three is the same. This also appears
to hold for the staggered action~\cite{Constantinou:2016ieh,Lee:1999zxa}.
We remark that, instead of extrapolating to the $N_f=3$ massless case,
we use physical quark
masses, which may be problematic, in particular regarding the strange
quark mass. However, in Ref.~\cite{Gockeler:2010yr} it was demonstrated
that the effect of the mass-dependence is tiny for the
perturbative momentum transfers that we are interested in. Moreover,
the difference is expected to vanish after a continuum
limit extrapolation of a renormalized matrix element
is carried out because our quark masses are tuned
to a line of constant physics.

Since the spin degrees of freedom are spread over hypercubes for staggered
fermions, the determination of the vertex function in momentum space is
more challenging than for Wilson fermions. We follow the approach described
in Ref.~\cite{Lytle:2013qoa}: the taste and spin degrees of freedom are
reconstructed from different momentum combinations. The quark propagator
for a given momentum, as any vertex function, will be a matrix of
size $16 \times 16$, after averaging over the color degrees of freedom.
Our choice of the scalar and tensor currents, where, in the
latter case, we employ a two-link operator, is detailed in
Ref.~\cite{Bali:2012jv}.

\begin{table}[bh]
\centering
  \begin{tabular}{|c|ccc|}
    \hline
    $\beta$&$a$/fm&$Z_T$&$Z_S$\\
    \hline
    3.45&0.282&1.07(12) &1.14(17) \\
    3.55&0.217&1.114(45)&0.829(12)\\
    3.67&0.153&1.125(19)&0.788(41)\\
    3.75&0.125&1.123(19)&0.723(38)\\
    3.85&0.099&1.100(18)&0.660(34)\\
    \hline
  \end{tabular}
  \caption{\label{zts}Conversion factors to the $\overline{\rm MS}$
    scheme at $\muqcd=2\textmd{ GeV}$.}
\end{table}
\begin{figure}[t]
 \centering
 \includegraphics[width=8.5cm]{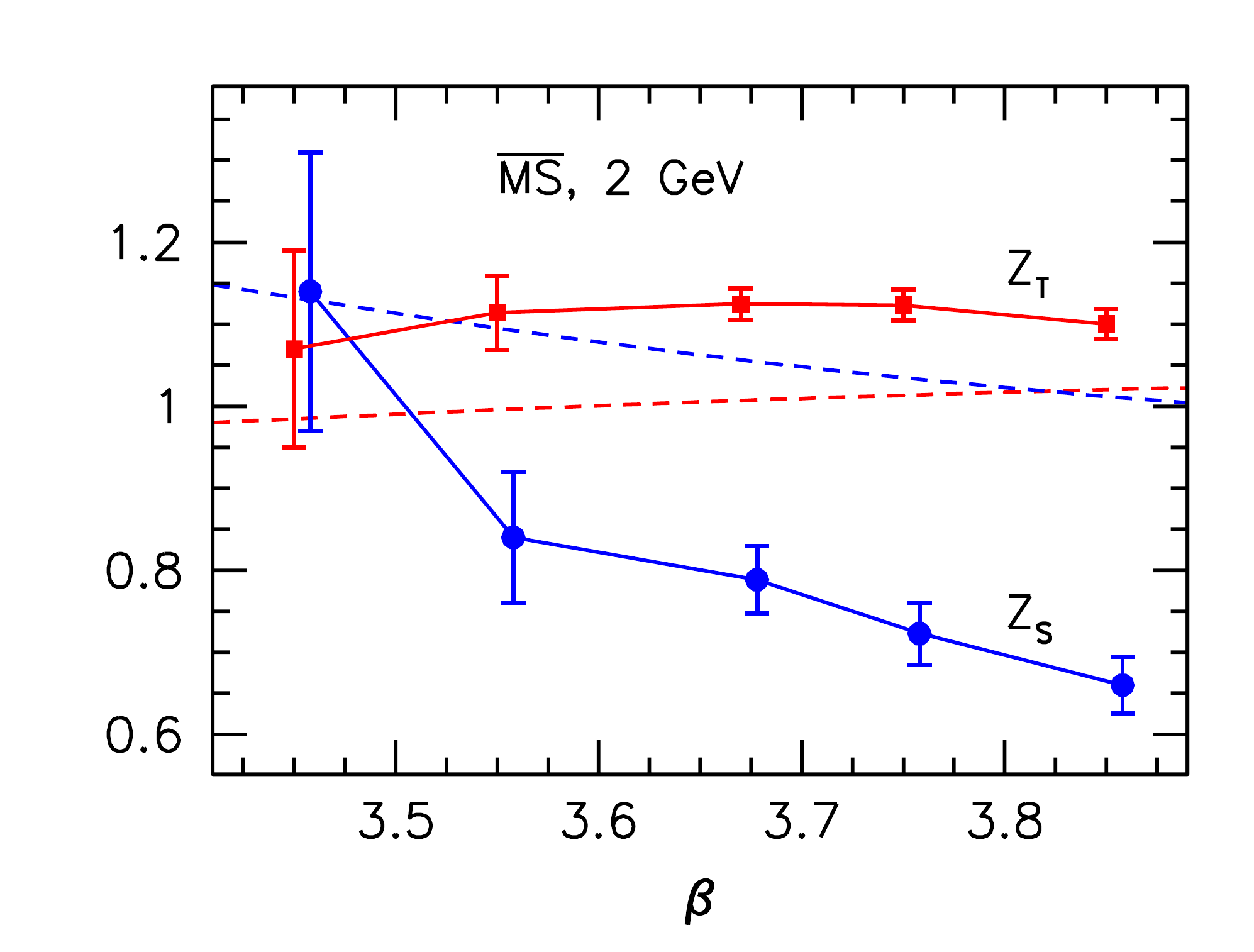}
 \caption{\label{fig:renconsts}Multiplicative renormalization constants
   as a function of $\beta$. The symbols have been slightly shifted
   horizontally for better visibility and connected by lines to guide the
   eye. Also shown as dashed lines are the one-loop perturbative
   expectations~\protect\cite{Bali:2012jv} that will be approached
   as $\beta\rightarrow\infty$.}
\end{figure}

The error of the final renormalization constants is dominated by
systematics. On the one hand, the conversion factors from the RI'-MOM
to the $\overline{\textrm{MS}}$-scheme are only known up to a fixed order
in perturbation theory (three loops in our case). Hence high momenta
are preferable. On the other hand, at momentum scales close to the
lattice cut-off the intermediate matching to the RI'-MOM scheme will
significantly be affected by lattice artifacts. Therefore,
we are restricted to a ``window'' of intermediate momentum values.
We employ combinations along symmetric lattice
directions, where the lattice corrections are smallest.
Another complication is that due to the choice of the
staggered action, the maximum momentum scale that can be achieved
on a four-dimensional lattice
is $\pi/a$, rather than $2\pi/a$. As a compromise, on the finest three lattices
we interpolate the RI'-MOM result to the fixed scale
$\muqcd=2\textmd{ GeV}$. Subsequently, this is perturbatively
converted to the $\overline{\rm MS}$-scheme. We estimate the uncertainty
by adding the difference between the scheme conversion at two- and
at three-loop order and the (statistical and systematic) interpolation
uncertainty in quadrature. The latter contribution is negligible.
At the coarsest two lattice spacings, $\muqcd=2\textmd{ GeV}$
is too close to the cut-off scale to obtain reliable results. Therefore,
at $a\approx 0.28\textmd{ fm}$ and at $a\approx 0.22\textmd{ fm}$,
we convert the RI'-MOM results at $\muqcd=1.1\textmd{ GeV}$ and
at $\muqcd=1.5\textmd{ GeV}$, respectively, to the $\overline{\rm MS}$-scheme
and evolve the result to $\muqcd=2\textmd{ GeV}$.
We replicate the same procedure at $a\approx 0.15\textmd{ fm}$
and add the difference that we obtain at this lattice spacing between
the matching at $\muqcd=2\textmd{ GeV}$ and the matching at these
lower scales in quadrature to the systematic error
at $\muqcd=1.1\textmd{ GeV}$ and $\muqcd=1.5\textmd{ GeV}$.

The results are listed in Table~\ref{zts} and shown in
Fig.~\ref{fig:renconsts}. We also include the
lattice perturbative theory one-loop expectations~\cite{Bali:2012jv}
in the figure. The comparatively larger value of $Z_T$ results
in a larger modulus of the renormalized tensor coefficient.

\section{Susceptibilities via current-current correlators}
\label{app:derivatives}

Here we derive Eqs.~\eqref{eq:chibderiv1} and~\eqref{eq:Hdef_tau} of 
the main text. To this end we consider a background field that 
possesses nonzero momentum $p_1$ in the $x_1$ direction. The constant field setup will be 
approached via the $p_1\to0$ limit. 
This approach has been described in detail in Ref.~\cite{Bali:2015msa} for $\chi_b$ 
in momentum space. Here 
we repeat the argument in coordinate space and also generalize it for $\tau_{fb}$. 

\subsection{Magnetic susceptibility from correlators}

We consider an oscillatory magnetic field and the corresponding Landau-gauge vector potential,
\be
B(x_1) = B \cdot \cos(p_1x_1), \quad\quad A_2(x_1) = B\cdot \frac{\sin(p_1x_1)}{p_1}\,.
\label{eq:oscBfield1}
\ee
The latter couples to $i\cdot e$ times the current~\eqref{eq:current} in the action density.
We can define the associated susceptibility just like in Eq.~\eqref{eq:defchi},
\be
\chi_b^{p_1,{\rm cos}} = -\left.\frac{\partial^2 f}{\partial (eB)^2}\right|_{B=0} 
=-\frac{T}{V}\int \dd^4 y\, \dd^4 z \,\frac{\sin (p_1y_1)}{p_1} \frac{\sin (p_1z_1)}{p_1}
\expv{j_2(y) j_2(z)}\,,
\ee
where each derivative brought down an integral over the current $j_2$ times the coordinate-dependence 
of $A_2$ and we used $\expv{j_2}=0$.
Changing the integration variable from $z$ to $x=z-y$ and exploiting the translational invariance of the current-current correlator, the integrals over $y_2$, $y_3$ and $y_4$ can be carried out,
\be
\chi_b^{p_1,{\rm cos}} = 
-\frac{1}{L} \int \dd y_1 \, \dd x_1 \,\frac{\sin(p_1 y_1)\sin(p_1(y_1+x_1))}{p_1^2}\, G(x_1)\,,
\label{eq:oscsusc1}
\ee
where the projected correlator $G(x_1)$, defined in Eq.~\eqref{eq:G22def}, appears.
In the $p_1\to0$ limit, $B(x_1)$ becomes homogeneous and $\chi_b^{p_1,{\rm cos}}$ equals the ordinary susceptibility $\chi_b$. 

For reasons that will become clear in a moment, let us consider a different background field,
\be
B(x_1) = B\cdot \sin(p_1 x_1), \quad\quad A_2(x_1) = -B\cdot \frac{\cos(p_1x_1)}{p_1}\,,
\label{eq:oscBfield2}
\ee
for which the associated oscillatory susceptibility, similarly to Eq.~\eqref{eq:oscsusc1}, reads
\be
\chi_b^{p_1,{\rm sin}} = -\frac{1}{L}\int \dd y_1\, \dd x_1 \,\frac{\cos (p_1y_1)\cos(p_1(y_1+x_1))}{p_1^2}\,G(x_1)
\,.
\ee
In this case the $p_1\to0$ limit does not reproduce $\chi_b$. Instead, $A_2(x_1)$ becomes 
homogeneous: it acts as if we had introduced a constant imaginary `chemical potential'
in the $x_2$ direction, with magnitude $\mu_2=-eB/p_1$. Therefore the oscillatory susceptibility
becomes proportional to the leading response to this spatial chemical potential,
\be
\chi_b^{p_1,{\rm sin}}\xrightarrow{p_1\to0} \frac{c_2}{p_1^2}, \quad\quad c_2 = -\frac{1}{L}\int \dd y_1 \,\dd x_1 \,
G(x_1)\,.
\ee

This detour was necessary to simplify the $p_1\to0$ limit of the oscillatory susceptibilities. Specifically, let us examine the following combination:
\begin{align}
&\chi_b^{p_1,{\rm cos}} + \chi_b^{p_1,{\rm sin}} - \frac{c_2}{p_1^2}\nonumber\\
&\qquad = 
-\frac{1}{L}\int \dd y_1 \,\dd x_1 \,\frac{ \sin (p_1y_1)\sin (p_1(y_1+x_1))
+ \cos (p_1y_1)\cos (p_1(y_1+x_1)) - 1 }{p_1^2}\,
G(x_1)\,. \label{eq:E7b}
\end{align}
This approaches $\chi_b$ for $p_1\to0$.
Using the trigonometric identity for the cosine of the difference of angles in the numerator of the kernel reveals that the integrand 
is independent of $y_1$. (This is why we needed to consider both the 
$\cos$- and $\sin$-type fields.)
Integrating over $y_1$ cancels the prefactor $1/L$, 
resulting in
\be
\chi_b =- \lim_{p_1\to0}
\int \dd x_1\, \frac{\cos(p_1x_1)-1}{p_1^2}\, G(x_1) 
= \int \dd x_1\, \frac{x_1^2}{2}\, G(x_1) \,,
\label{eq:chibderiv2}
\ee
where we finally performed the $p_1\to0$ limit. This 
proves Eq.~\eqref{eq:chibderiv1} of the main text. 
We note that the crucial point of the derivation was Eq.~\eqref{eq:E7b}, 
where the kernel was shown to only depend on the distance $x_1$ 
between the two current insertions. This was done conveniently 
using the combination of the oscillatory fields -- in contrast, it 
would have been more tedious if we started directly with a constant 
background.

\subsection{Tensor coefficient from correlators}

We generalize the above derivation for $\tau_{fb}$, which can be written as
\be
\tau_{fb} = \frac{1}{q_f/e}  \left.\frac{\partial }{\partial (eB)}\right|_{B=0} \frac{T}{V} \int \dd^4 x\,\expv{\bar\psi_f \sigma_{12} \psi_f(x)}\,.
\ee
Again we consider oscillatory magnetic fields of the types~\eqref{eq:oscBfield1} and~\eqref{eq:oscBfield2}. These give rise to modulated tensor bilinears of the 
forms $\bar\psi_f \sigma_{12} \psi_f(x) \cos(p_1x_1)$ and $\bar\psi_f \sigma_{12} \psi_f(x) \sin(p_1 x_1)$, respectively,
which enter the corresponding oscillatory tensor coefficients $\tau_{fb}^{p_1,{\rm cos}}$ and $\tau_{fb}^{p_1,{\rm sin}}$:
\be
\begin{split}
\tau_{fb}^{p_1,{\rm cos}} &= 
\frac{i}{q_f/e}\,\frac{1}{L}\int \dd y_1\, \dd x_1 \,\cos (p_1y_1)\, \frac{\sin (p_1(y_1+x_1))}{p_1}
\,H_f(x_1)\,,\\
\tau_{fb}^{p_1,{\rm sin}} &= 
\frac{-i}{q_f/e}\,\frac{1}{L}\int \dd y_1\, \dd x_1 \,\sin (p_1y_1) \,\frac{\cos (p_1(y_1+x_1))}{p_1}
\,H_f(x_1)\,,
\end{split}
\ee
where the projected tensor-vector correlator $H_f(x_1)$, defined in Eq.~\eqref{eq:Hdef_tau},
appears. Here we performed the same variable substitution as in Eq.~\eqref{eq:oscsusc1} above.

\begin{figure}[t]
 \centering
 \includegraphics[width=8.5cm]{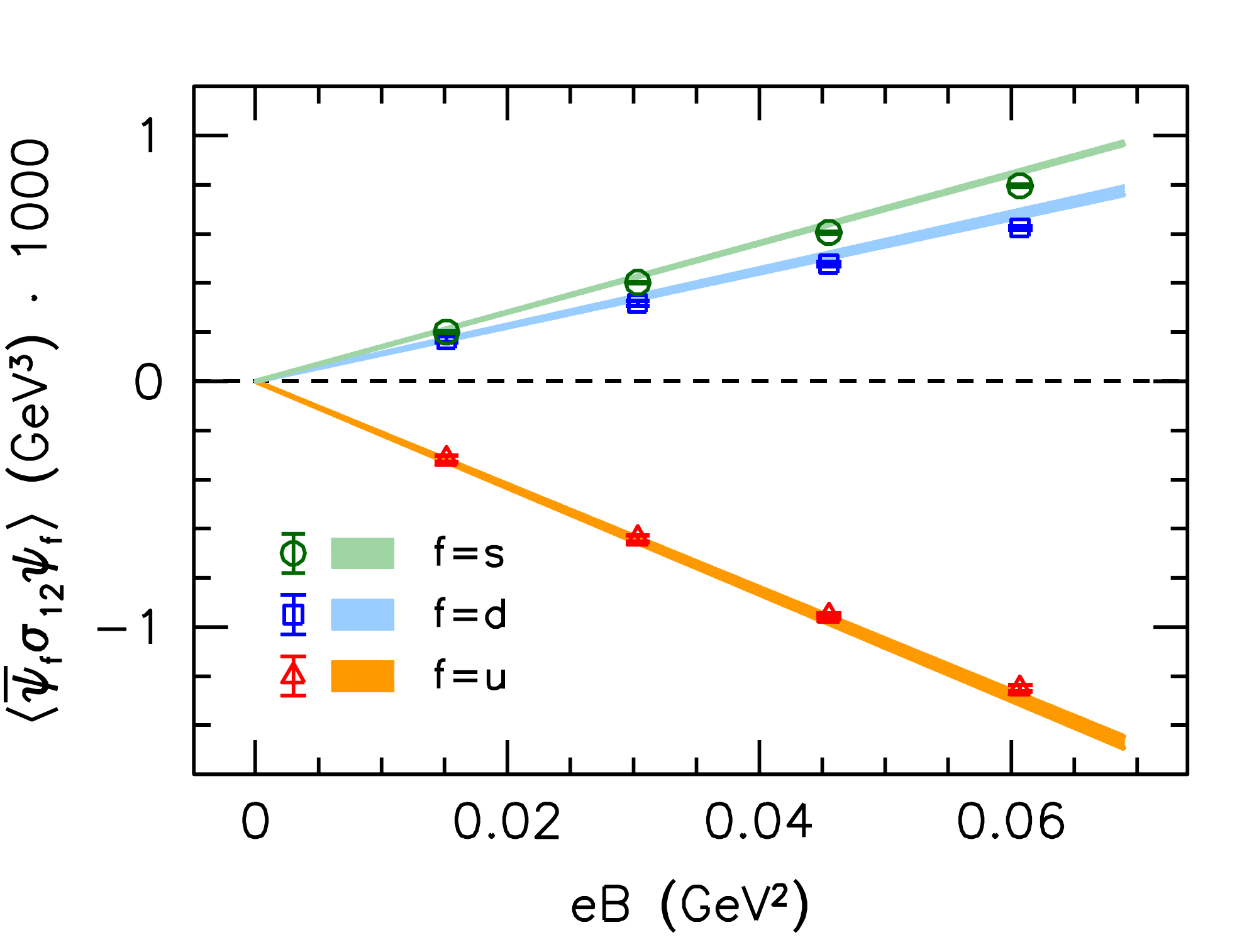}
 \caption{\label{fig:tau_as_der}
 Comparison of different methods to calculate $\tau_{fb}$ for 
all three flavors. 
Simulations at nonzero (quantized) values of the magnetic field
(points) are compared with a direct determination of the slope at $B=0$
(colored bands).
}
\end{figure}

In this case, $\tau_{fb}^{p_1,{\rm cos}}$ approaches $\tau_{fb}$ for $p_1\to0$, while $\tau_{fb}^{p_1,{\rm sin}}$ vanishes in that limit. 
Thus we need to consider the sum of the two coefficients. Employing the trigonometric
identity for the sine of the difference of angles and carrying out the integral over $y_1$ gives
\be
\tau_{fb}=\lim_{p_1\to0} \left[
\tau_{fb}^{p_1,{\rm cos}} + \tau_{fb}^{p_1,{\rm sin}} \right]
= 
\lim_{p_1\to0}
\frac{i}{q_f/e}\,\int \dd x_1 \,\frac{\sin (p_1x_1)}{p_1}\,
H_f(x_1)
=
\frac{i}{q_f/e} \int \dd x_1\, x_1 \,H_f(x_1)\,.
\ee
In finite volumes we carry out
the same symmetrization as in Eq.~\eqref{eq:Pi0def}, 
this time taking into account that $H_f(x_1)=-H_f(L-x_1)$ to
finally arrive at Eq.~\eqref{eq:Hdef_tau} of the main text.

This method to calculate $\tau_{fb}$ is compared to 
the results for $\expv{\bar\psi_f\sigma_\subs\psi_f}$ measured at $B>0$
on $24^3\times6$ lattices at $T=113\textmd{ MeV}$ 
in Fig.~\ref{fig:tau_as_der}.
For the light quarks we obtain consistent results, however, for $\tau_{sb}$ the 
correlator tends to give values that slightly differ from the
slope of a linear fit to the lowest few 
available points. Since lattice artifacts and finite volume effects might be different 
in the two cases, such slight differences are not unexpected.

In addition, we find that a linear fit to results from simulations
at $B>0$ has smaller uncertainties than extracting the slope at $B=0$
using the correlator method. In the main text we therefore use our earlier 
results for $\expv{\bar\psi_f\sigma_\subs\psi_f}$ from Ref.~\cite{Bali:2012jv}.

We note that the tensor-vector correlators at nonzero spatial momenta 
might also be useful for extracting further features of the photon
distribution amplitude.

\section{Parametrization of the equation of state}
\label{app:parameterization}

Up to $\mathcal{O}(B^2)$, the magnetic field-dependence of the 
complete EoS can be calculated from the magnetic susceptibility $\chi(T)$. Here we provide a parametrization for this 
observable and also collect the relevant
thermodynamical relations, which were also summarized in Ref.~\cite{Bali:2014kia}.

First of all, we remind the reader that in the presence of a background magnetic field, the different components of the pressure -- defined by considering
an infinitesimal compression of the system in the respective direction -- 
might become anisotropic~\cite{Bali:2013esa}. 
In particular, one should distinguish between the \emph{$\Phi$-scheme}, where the flux of the magnetic field is kept constant during the compression (superscript $(\Phi)$ below), and the \emph{$B$-scheme}, where the magnetic field strength is kept constant (superscript $(B)$). 
On the one hand, the $B$-scheme pressure is isotropic and equals the negative of 
the free energy density in the thermodynamic limit,
\be
 p_{1,2}^{(B)} =p_3 = -f\,.
 \label{eq:f_p3}
\ee
On the other hand, in the $\Phi$-scheme the pressure components 
are related by 
the magnetization $\M$,
\be
p_{1,2}^{(\Phi)} = p_{3} - eB\cdot \M, \quad\quad \M = -\frac{\partial f}{\partial (eB)}\,.
\ee

The entropy density $s$ and the energy density $\epsilon$ are scheme-independent,
\be
s = -\frac{\partial f}{\partial T}, \quad\quad \epsilon = f + Ts\,,
\ee
whereas also the interaction measure (trace anomaly) $I$
differs between the two schemes,
\be
I^{(B)} = \epsilon - 3 p_3, \quad\quad 
I^{(\Phi)} = \epsilon - 2p_{1,2}^{(\Phi)} - p_3 = I^{(B)}+2 eB\cdot \M\,.
\label{eq:I_B_Phi}
\ee
Using Eqs.~\eqref{eq:defchi} and~\eqref{eq:f_p3}, the leading-order expansion in the magnetic field takes the form
\be
p_3(T,B) = p_3(T,0) +\chi(T) \,\frac{(eB)^2}{2}, \quad\quad 
\M(T,B) = \chi(T)\, eB\,.
\ee
Together with Eqs.~\eqref{eq:f_p3}--\eqref{eq:I_B_Phi} these specify the 
$B$-dependence of all relevant observables up to $\mathcal{O}(B^2)$.

At $B=0$ the pressure is isotropic, and can be 
obtained from the interaction measure as\footnote{We note that Eq.~\eqref{eq:pfromI}) remains valid also for $B>0$ in the $B$-scheme 
but not in the $\Phi$-scheme.}
\be
\frac{p(T,B=0)}{T^4} = \int_0^{T'} \!\!\dd T' \,\frac{I(T',B=0)}{T^5}\,.
\label{eq:pfromI}
\ee
Thus, to calculate the complete EoS including $B^0$ and $B^2$ effects, 
altogether it suffices to parameterize $I(T,0)$ and $\chi(T)$. 
For the latter we consider a parametrization of the continuum extrapolated lattice results that smoothly approach the 
HRG model prediction (see Fig.~\ref{fig:chiT}) at low and 
the perturbation theory formula~\eqref{eq:highTchi}) at high temperatures.
We found the following 
parametric form to be sufficient for this,
\be
\chi(T) = \exp(-h_3/t)\cdot \frac{1+g_0/t + g_1/t^2 + g_2/t^3}{1+ g_3/t + g_4/t^2 + g_5/t^3} \cdot 2\beta_1 \log\frac{t}{q_0}\,, \quad\quad 
t=\frac{T}{1 \textmd{ GeV}}\,.
\label{eq:chipars}
\ee
Eq.~\eqref{eq:chipars} incorporates the non-perturbative temperature-dependence predicted by the HRG model (see App.~\ref{sec:B}) at low $T$
and the logarithmic rise at high temperatures. The $\beta_1$ 
coefficient is fixed to its perturbative value~\eqref{eq:beta1def}, 
while the scale $q_0$ inside the logarithm is allowed to be a free 
parameter.
The rational function
involving the $g_i$ parameters interpolates between the two 
limiting behaviors.
The so-obtained parametrization is shown in the left 
panel of Fig.~\ref{fig:artefacts}
in the main text.

For the interaction measure we take the parametrization of Ref.~\cite{Borsanyi:2013bia},
     \be
\frac{I(T,0)}{T^4} = \exp(-h_1/t-h_2/t^2)\cdot \left(h_0 + \frac{f_0\cdot\left[ \tanh(f_1\cdot t + f_2) + 1 \right]}{1+k_1\cdot t + k_2\cdot t^2} \right)\,, 
\quad\quad t=\frac{T}{0.2 \textmd{ GeV}}\,.
\label{eq:Ipars}
\ee
The parameters of both functions are included in Table.~\ref{tab:pars}.
The two parametrizations, together with the implementations of the
formulae~\eqref{eq:f_p3}--\eqref{eq:pfromI} are included in the Python script
{\tt param\string_EoS.py} that is submitted to arXiv.org together with this 
manuscript.

\begin{table}[ht!]
 \centering
 \begin{tabular}{|c|c|c|c|c|c|c|c|c|}
 \hline
 $\beta_1$ & $h_3$ & $g_0$ & $g_1$ & $g_2$ & $g_3$ & $g_4$ & $g_5$ & $q_0$ \\ \hline
  $1/(6\pi^2)$ & 0.1544 & 23.99 & -2.085 & 0.1290 & 21.35 & -6.201 & 0.5766 & 0.1497 \\ \hline
 \end{tabular}
 \begin{tabular}{|c|c|c|c|c|c|c|c|}
 \hline
  $h_0$ & $h_1$ & $h_2$ & $f_0$ & $f_1$ & $f_2$ & $k_1$ & $k_2$ \\ \hline
  0.1396 & -0.1800 & 0.0350 & 1.05 & 6.39 & -4.72 & -0.92 & 0.57 \\ \hline
 \end{tabular}
 \caption{\label{tab:pars}Parameters of the functions~\protect\eqref{eq:chipars} and~\protect\eqref{eq:Ipars}.}
\end{table}

This parametrization is valid for low magnetic fields. 
To be more quantitative, we compare our $\mathcal{O}(B^2)$ truncated 
results for the longitudinal pressure 
to the complete magnetic field-dependence from Ref.~\cite{Bali:2014kia} for $T\gtrsim180 \textmd{ MeV}$. 
We find agreement within errors in the range $B/(\pi T)^2 \lesssim 1$. This upper 
limit is hard-coded in the Python script as well.
One final remark about the parametrization is in order. All truncated 
thermodynamic observables approach zero for $T\to0$, such that a normalization 
by the corresponding powers of the temperature (i.e.\ $p_3/T^4$, $s/T^3$ and so on) 
produces sensible plots. This is not the case if $\mathcal{O}(B^4)$ terms are also 
included: at this order vacuum contributions arise and the equation of state depends 
on $B$ already at $T=0$, rendering a normalization like $p_3/T^4$ ill-defined 
in the $T\to0$ limit~\cite{Bali:2014kia}.
\newline

\bibliographystyle{jhep_new}
\bibliography{suscep}
\end{document}